\documentclass[reprint,amsmath,amssymb,aps]{revtex4-2}
\usepackage{graphicx}
\usepackage{dcolumn}
\usepackage{bm}
\usepackage{braket}
\usepackage{hyperref}       
\usepackage{url}
\usepackage{multirow}

\begin{document}

\title{Curvature-Aware Deep Learning for Vector Boson Fusion: Differential Geometry, Physics-Inspired Features, and Quantum Method Limitations }
\author{Alibordi Muhammad}
\email{alibordi.muhammad@cern.ch}
\affiliation{Faculty of Physics, University of Warsaw, Ludwika Pasteura 5, 02-093 , Warsaw, Poland}

\date{\today}

\begin{abstract}
Particle physics classification traditionally embeds collision events in flat Euclidean spaces, discarding geometric information encoded in curved statistical manifolds. This work develops a differential geometric framework for Vector Boson Fusion Higgs classification, integrating product manifold neural networks with physics-inspired feature engineering. Five observables capturing momentum correlations, angular interference, mass superposition, topology, and hierarchical structure translate quantum field theory concepts into classical features. Product manifold architectures decomposing representations across Euclidean, hyperbolic ($\kappa=-1.5$), and spherical ($\kappa =+1.0$) geometries achieve 0.9454 AUC, representing 0.31 percent improvement over Euclidean baselines on one million events. Physics-inspired features contribute an additional 0.24 percent when geometrically embedded, totaling 0.57 percent gain—meaningful for rare signals at 40:1 background ratio. Individual features show weak standalone discrimination but become effective through manifold alignment, validating that geometric scaffolding enables multivariate correlations. Quantum kernel methods, while theoretically interesting, achieve only 0.667 AUC on small subsets, demonstrating current computational intractability. Three principles emerge: geometric structure provides necessary improvements when Fisher-Rao curvature is significant, physics-inspired features require manifold scaffolding for effectiveness, and generalization scales polynomially with curvature bounds under classical computation.
\end{abstract}

\maketitle

\section{Introduction}
\label{sec:intro}
\noindent  The central idea is that curvature awareness in a machine learning model allows the capture of non-linear correlations among input features of hierarchical data~\cite{ig1,ig2,ig4}. Standard models of machine learning often rely on the Euclidean assumption, where features are treated as independent coordinates embedded in $\mathbb{R}^n$ and distances are measured through the flat line element $ds^2 = \sum_{i=1}^n (dx^i)^2$. This assumption implicitly neglects intrinsic dependencies among features, thereby discarding relevant structural information. In reality, the correlations are often non-linear and intrinsic to the data, and a more faithful description requires endowing the feature space with a curved geometric structure. In such a setting, distances are instead measured as $ds^2 = g_{ij}(\bm{x}) \, dx^i dx^j,$ where $g_{ij}(\bm{x})$ denotes the local geometry induced, for example, by the Fisher information matrix~\cite{ig1,qfi1,qfi2} or other information-geometric constructions. The learning task of a machine learning model can then be viewed as estimating the posterior probability $p(\mathcal{H} \mid \bm{x}),$ with $\bm{x}$ denoting the feature vector and $\mathcal{H}$ the hypothesis space, while respecting the curved structure of the statistical manifold. Curvature-aware models therefore transcend the Euclidean paradigm by encoding correlations that are intrinsic to the data rather than imposed by linear embeddings, thus ensuring that the representation of hierarchical dependencies remains geometrically consistent.

\section{Vector Boson Fusion Reaction}
\label{subsec:vbfground}

\noindent The conceptual foundation of curvature-aware learning treats data as points on a curved statistical manifold rather than in a flat Euclidean space. To demonstrate the efficacy of this geometric framework, the Vector Boson Fusion (VBF) production of the Higgs boson serves as a concrete test case in high-energy physics. This process represents a well-studied yet algorithmically demanding benchmark whose multi-scale correlations (Fig.~\ref{fig:corr}) and interference patterns closely mirror the complexities expected in rare radiative decays such as $B_s^0 \to \mu^+\mu^-\gamma$. The choice of VBF as a test case is therefore strategic: it offers abundant data, a rich interplay of electroweak and QCD effects, and a complex event topology that stresses both the representation power and geometric sensitivity of the learning model~\cite{mf1,mf2,mf3}. Lessons drawn from VBF—particularly those concerning curvature-aware encoding of quantum correlations—are directly transferable to rare decays, where sparse statistics and non-linear feature dependencies pose even greater challenges.This article presents a novel framework for high-energy physics data analysis that bridges quantum field theory concepts with machine learning feature engineering. Five quantum-inspired features are developed specifically for Vector Boson Fusion (VBF) Higgs boson classification, each grounded in fundamental quantum mechanical principles and particle physics phenomenology. These features capture multi-scale correlations, quantum interference patterns, and field coherence effects that traditional kinematic variables cannot adequately represent.The VBF process, characterized by the reaction $pp \to H + 2j \to ZZ^* + 2j \to 4\ell + 2j$, produces a distinctive topology with two forward jets and four central leptons. However, this signature is contaminated by numerous background processes, including continuum $ZZ$ production, gluon fusion Higgs production, and various electroweak channels. Traditional kinematic variables such as transverse momentum ($p_T$), pseudorapidity ($\eta$), and invariant masses capture the basic physics of these processes but fail to exploit the deeper quantum mechanical correlations present in the data.The quantum field theory underlying particle interactions suggests that information about production mechanisms is encoded in subtle correlations between final state particles—correlations that mirror quantum entanglement, interference, and coherence phenomena. The theoretical foundation rests on the quantum–classical correspondence principle. While the detected particles are classical observables, their production process is governed by quantum field theory. The transition from quantum amplitudes to classical probabilities preserves certain correlation structures that can be recovered through appropriately designed observables. In quantum mechanics, entangled states exhibit correlations that cannot be explained by classical physics; analogously, the multi-body kinematic correlations in VBF encode non-trivial structure that can only be meaningfully decoded when the learning model respects the intrinsic geometry of the data manifold.

\begin{figure}[b]
\includegraphics[width=1.0\linewidth]{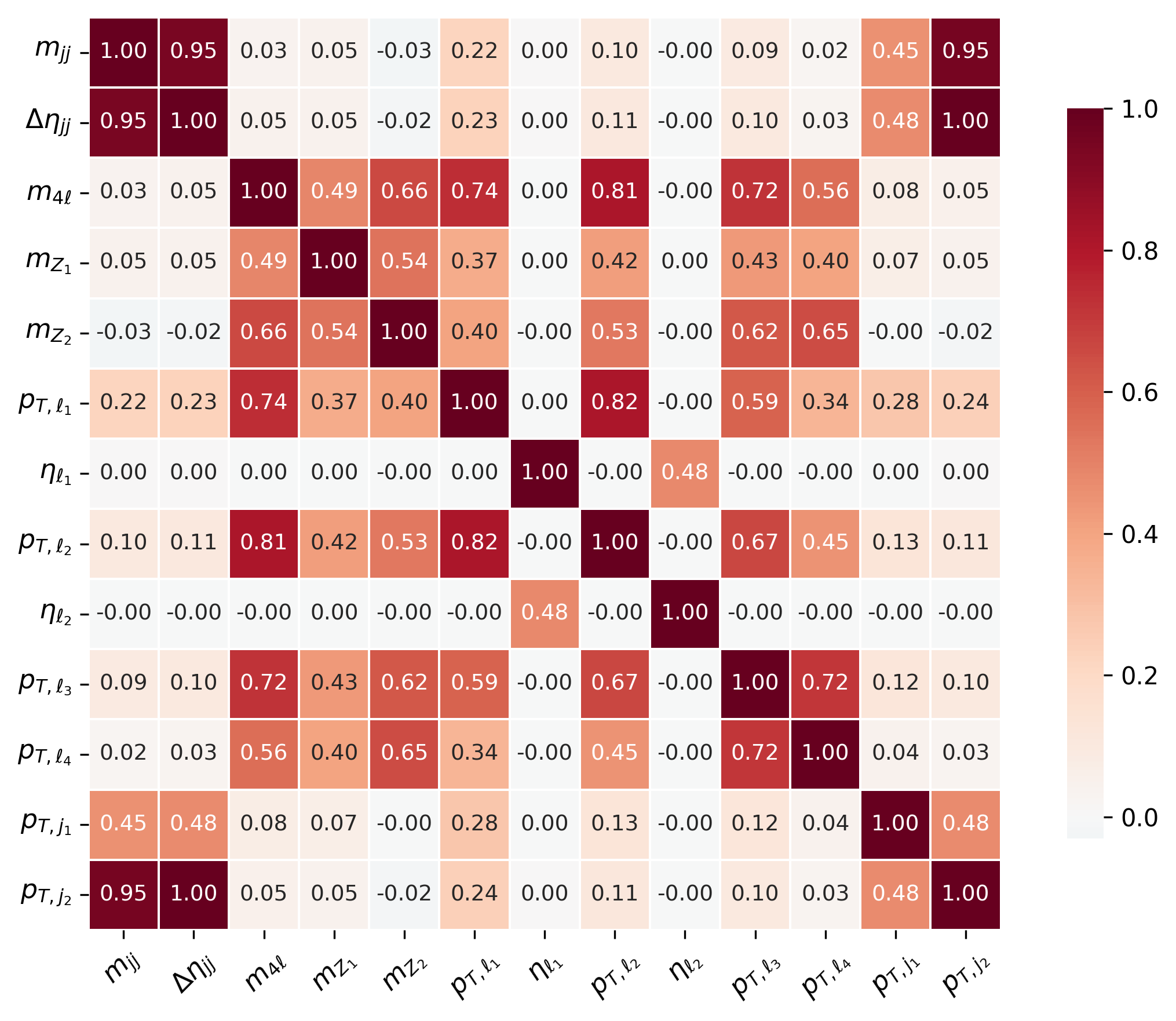}
\caption{ Classical correlation among the kinematic observables related to the Vector Boson Fusion decay process.}
\label{fig:corr}
\end{figure}

\subsection{\label{sec:level2}Event Representation}

\noindent The construction and representation of the event dataset used in this study treats each collision event as a point in a high-dimensional kinematic manifold~\cite{mf3,gdl1,gdl2} whose coordinates are defined by measurable four-momenta and derived observables. The dataset comprises both signal and background samples in the $\mathrm{H}\to ZZ\to4\ell$ decay channel. The signal hypothesis corresponds to VBF Higgs production, while the dominant background hypotheses include gluon–fusion Higgs production, diboson processes ($ZZ\to4\ell$), and associated production channels such as $WH\to ZZ\to4\ell$. All event samples are stored in \texttt{ROOT} format and organized in a reduced tree structure. A Python interface, implemented through \texttt{uproot}, reads these trees directly into \texttt{pandas} dataframes, enabling rapid vectorized data access. Each event is described by $N_{\text{dim}}=24$ kinematic observables, encompassing invariant masses $(m_{jj},\, m_{4\ell},\, m_{Z_1},\, m_{Z_2})$, dijet rapidity separation $(\Delta\eta_{jj})$, and momenta $(p_T,\, \eta,\, \phi)$ of four leptons and two leading jets. Metadata such as run number, event index, and generator weights are retained for bookkeeping but excluded from the feature space used for learning. Invalid numerical entries (NaN or $\pm\infty$) are filtered during ingestion. Events are assigned binary labels $y$, where $y=1$ corresponds to signal-like (VBF) events and $y=0$ to background-like events. When available, generator-level weights are used for statistical weighting; otherwise, all events receive unit weight. In cases where some input files are missing or corrupted, a synthetic generator produces statistically consistent surrogate data. This generator samples from parameterized distributions designed to emulate salient VBF characteristics—namely, large dijet invariant mass and wide rapidity gaps—as well as broader background topologies.
\begin{figure*}[t]
 \centering
 \includegraphics[width=0.32\linewidth]{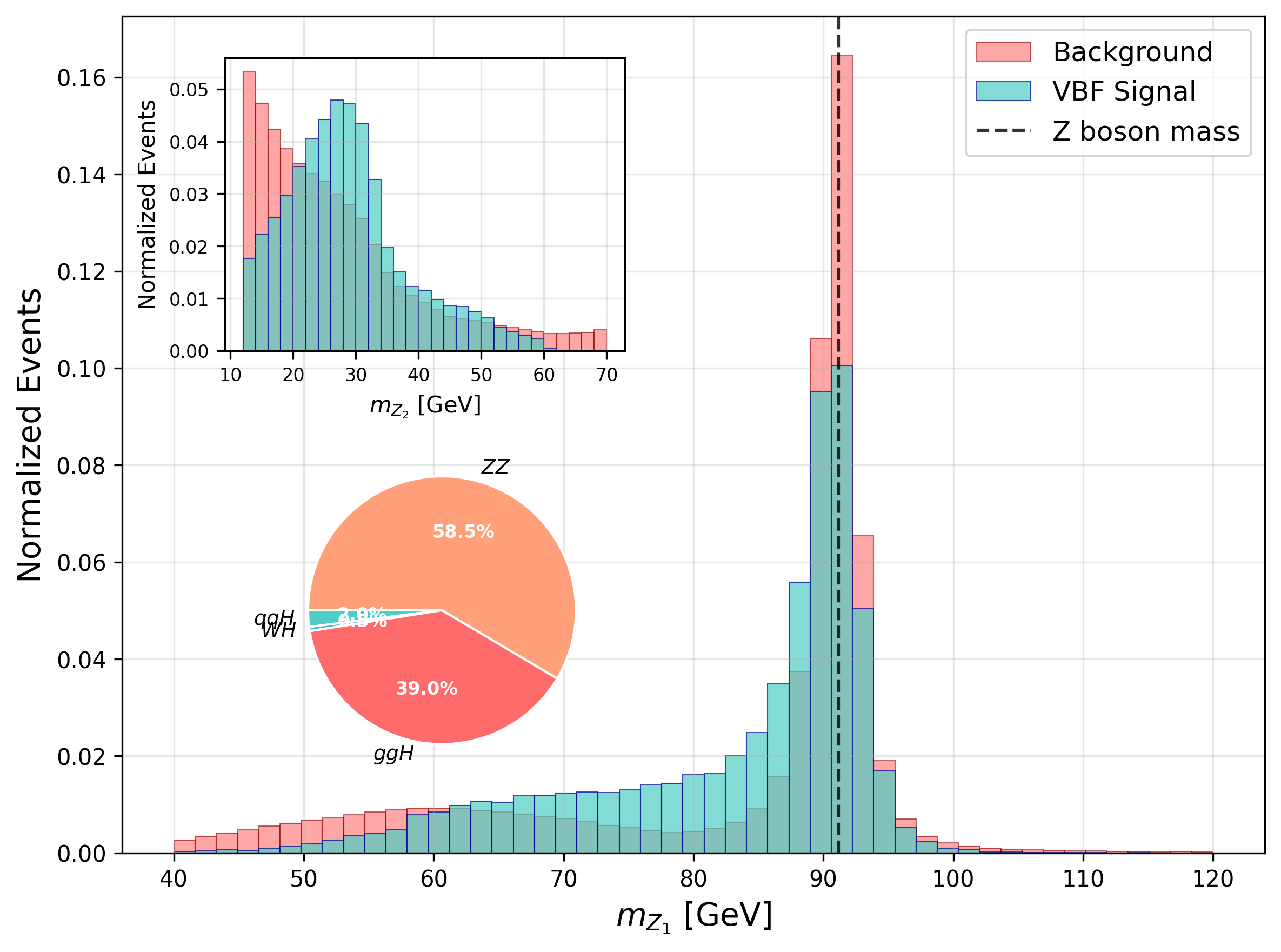}  \hfill
 \includegraphics[width=0.32\linewidth]{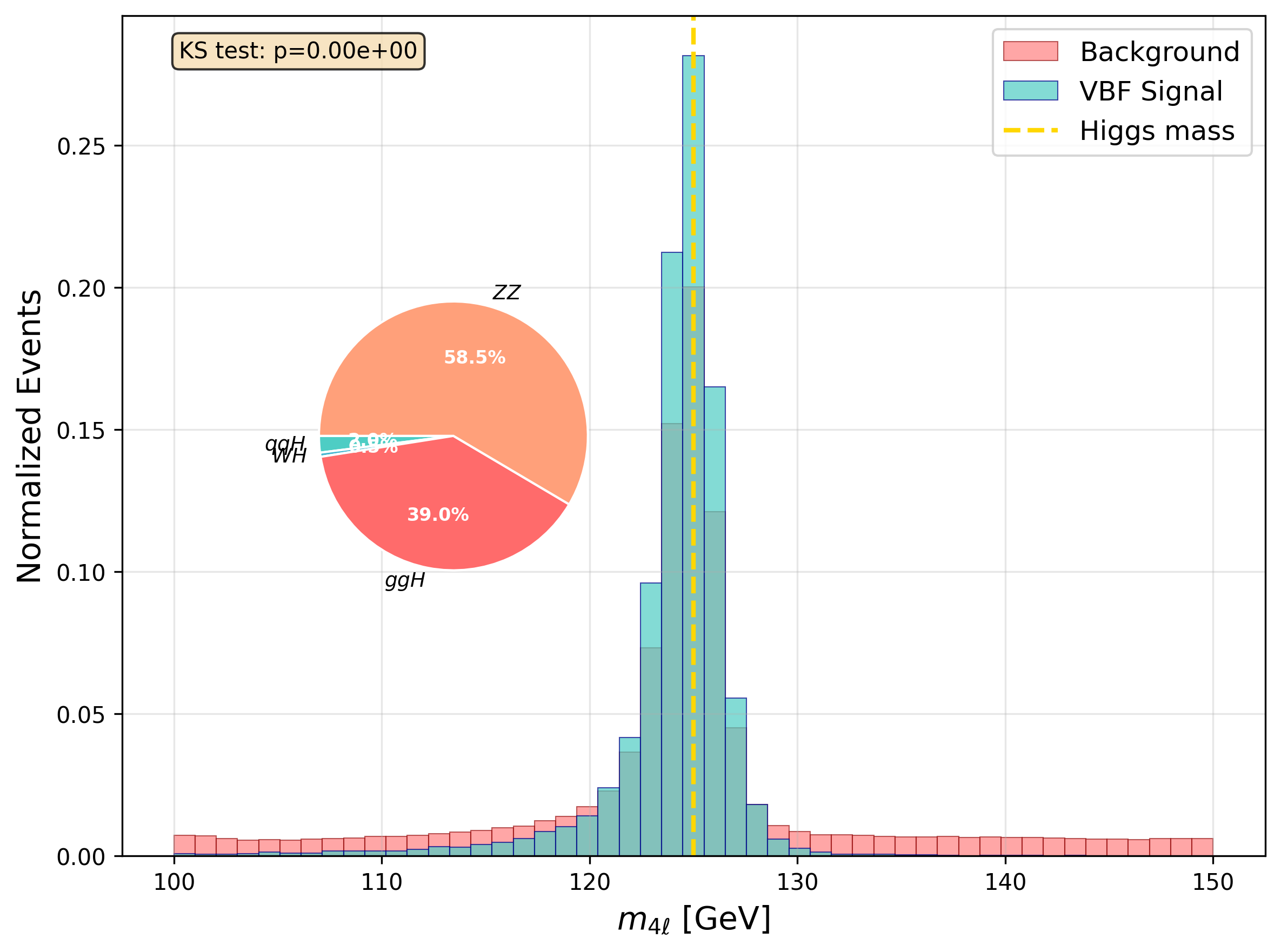}  \hfill
 \includegraphics[width=0.32\linewidth]{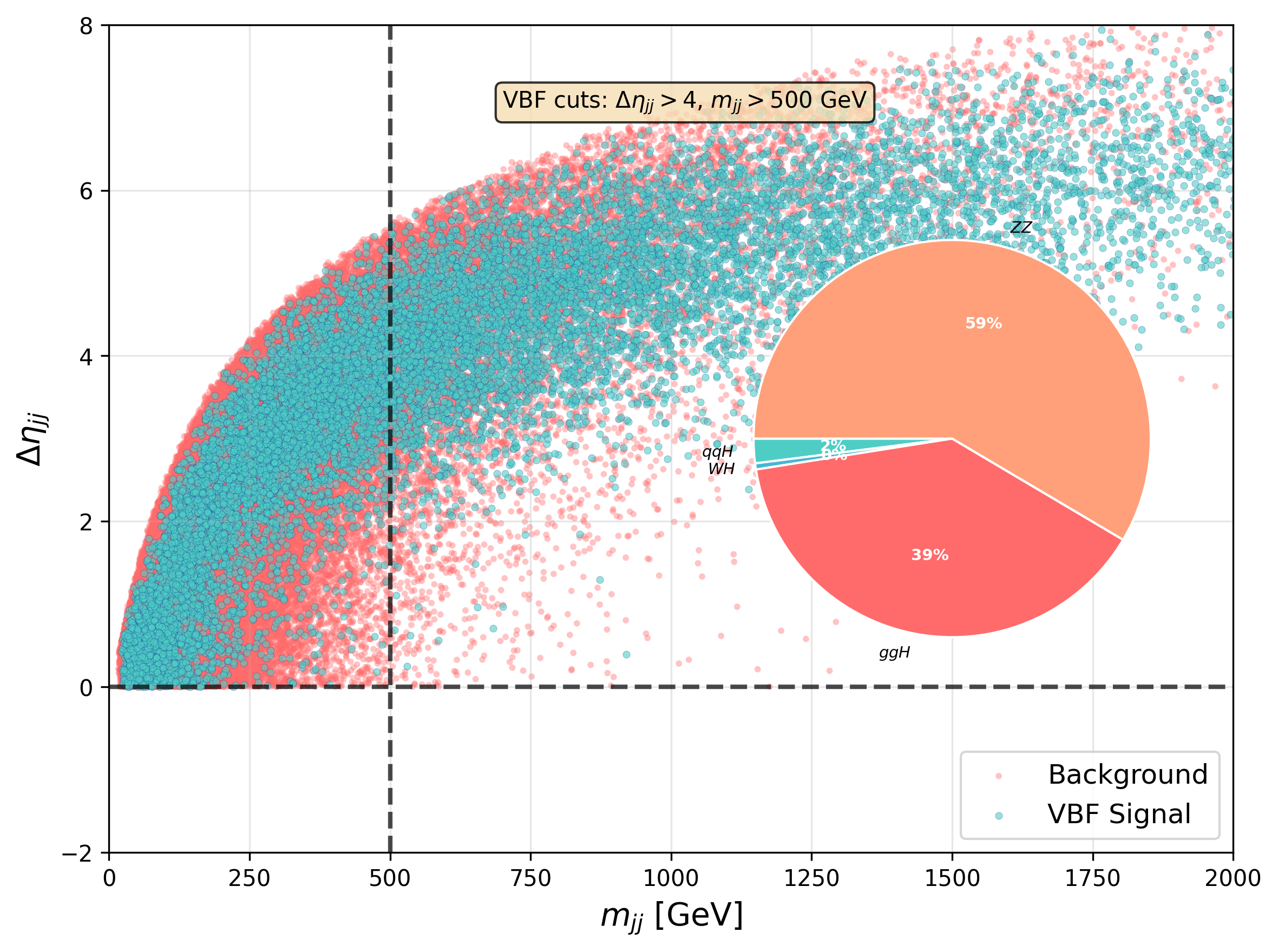}
  \caption{Representative kinematic distributions from the Drell–Yan sample. The panels illustrate, respectively, the Z$_1$ mass spectrum, the four-lepton invariant mass around the Higgs resonance, and the canonical VBF discriminating variables $m_{jj}$ and $\Delta\eta_{jj}$.}
 \label{fig:gendy}
\end{figure*}
The preprocessed dataset can thus be summarized as
\begin{equation}
X \in \mathbb{R}^{N\times D}, \qquad 
y \in \{0,1\}^N, \qquad 
w \in \mathbb{R}^N_{+},
\end{equation}
where $N$ is the total number of events and $D=23$ the number of retained kinematic features. After preprocessing, the dataset comprises nearly one million events across all categories, ensuring sufficient statistical power for both training and validation phases. The number of effective variables per event is $D=23$, capturing the dijet and four-lepton observables summarized in Table~\ref{tab:dataset}.
\begin{table}[h!]
\centering
\begin{tabular}{lrr}
\hline
Process & Events Loaded & Category \\
\hline
VBF $H\to ZZ \to 4\ell$ & 24,867 & Signal \\
ggH $H\to ZZ \to 4\ell$ & 134,682 & Background \\
$ZZ \to 4\ell$ & 817,660 & Background \\
$WH \to ZZ \to 4\ell$ & 20,934 & Background \\
\hline
Total & 998,143 & --- \\
\hline
\end{tabular}
\caption{Event statistics after preprocessing. The VBF $H\to ZZ \to 4\ell$ channel defines the signal class, while gluon–fusion Higgs, diboson, and associated production processes are grouped as backgrounds.}
\label{tab:dataset}
\end{table}
The event feature space is composed of dijet and four-lepton kinematics:
\begin{equation}
\{ f_{\text{massjj}},\, f_{\Delta jj},\, f_{\text{mass4}\ell},\,
f_{Z_1},\, f_{Z_2},\, f_{\ell_i}^{(p_T,\eta,\phi)},\,
f_{j_k}^{(p_T,\eta,\phi)} \},
\end{equation}
where $i=1,\dots,4$ indexes the leptons and $k=1,2$ the two leading jets. This yields 23 independent features encapsulating the essential geometric and kinematic degrees of freedom:
\begin{equation}
\mathcal{J}_i = \{(\mathbf{p}_{\ell_1}, \mathbf{p}_{\ell_2}, \mathbf{p}_{\ell_3}, \mathbf{p}_{\ell_4}), (\mathbf{p}_{j_1}, \mathbf{p}_{j_2}), m_{4\ell}, m_{jj}, \Delta\eta_{jj}\},
\end{equation}
where $\mathbf{p}_{\ell_j}=(p_{T,j},\eta_j,\phi_j)$ represents lepton momenta, and $\mathbf{p}_{j_k}$ the jet momenta. The global quantities $m_{4\ell}$, $m_{jj}$, and $\Delta\eta_{jj}$ capture the overall event topology. To visualize and validate the physical consistency of the input data, a suite of diagnostic plots is prepared. Each visualization combines statistical metrics with physics annotations to enhance interpretability. The first two panels in Fig.~\ref{fig:gendy} show the Z$_1$ and four-lepton mass spectra, while the third illustrates the separation power of VBF observables. The Z$_1$ invariant mass peaks around the $Z$ boson resonance but shows process-dependent tails. The four-lepton invariant mass $m_{4\ell}$, centered near $125$~GeV, is compared between signal and background distributions using the Kolmogorov–Smirnov statistic,
\begin{equation}
D_{\mathrm{KS}} = \sup_x |F_{\mathrm{sig}}(x) - F_{\mathrm{bkg}}(x)|,
\end{equation}
where $F(x)$ denotes the empirical cumulative distribution. 
\begin{figure}[h!]
  \centering
  \includegraphics[width=1.0\linewidth]{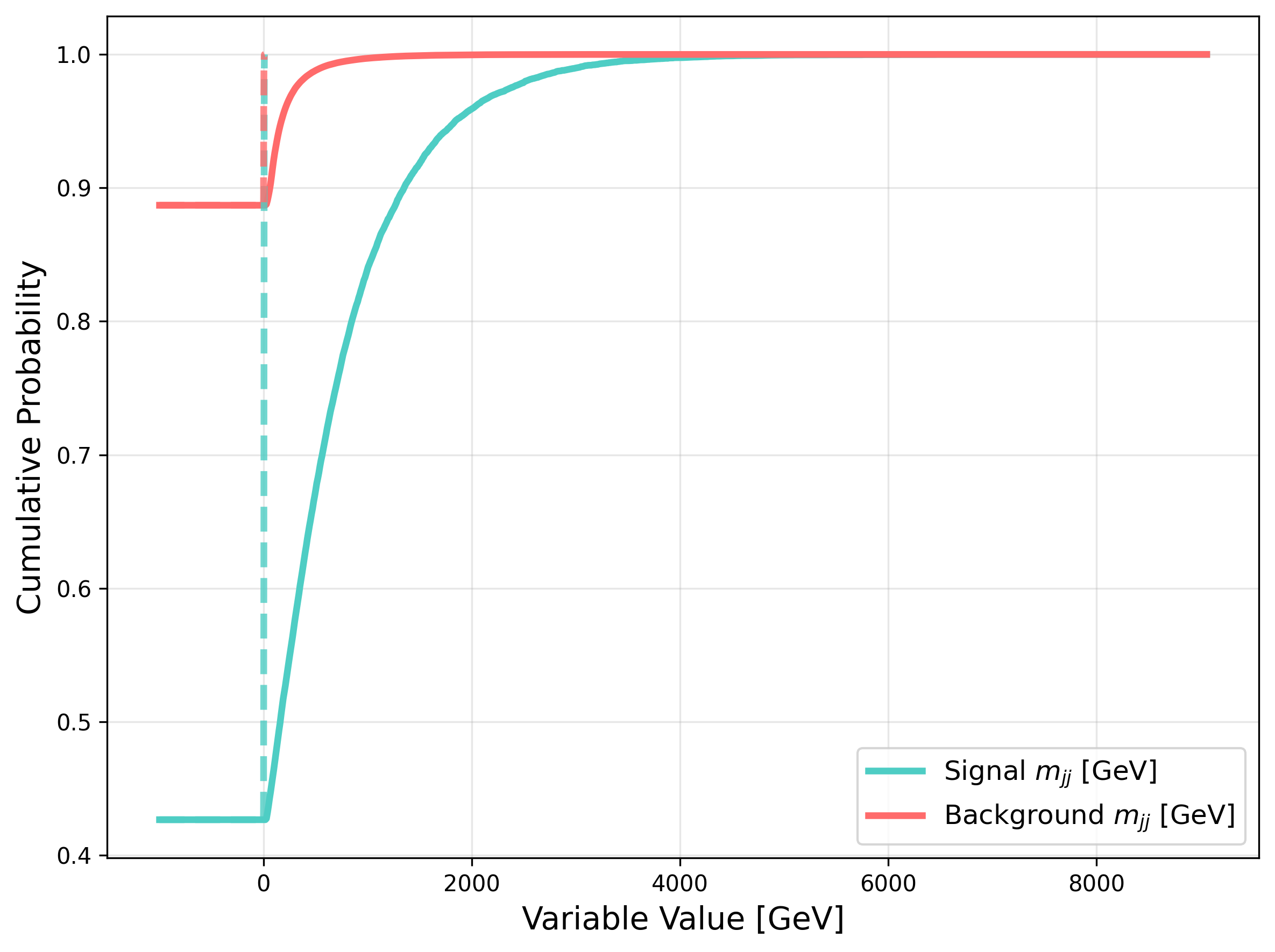}
  \caption{Cumulative distribution comparison for key observables highlighting tail behaviors between signal and background.}
  \label{fig:cdf}
\end{figure}
The canonical VBF discriminants—dijet invariant mass and rapidity gap—form a two-dimensional space in which signal enrichment is visible for $m_{jj}>500$~GeV and $\Delta\eta_{jj}>4$. For each observable, a scalar measure of discriminating power is defined as
\begin{equation}
S = \frac{|\mu_{\mathrm{sig}} - \mu_{\mathrm{bkg}}|}{\sqrt{\tfrac{1}{2}(\sigma_{\mathrm{sig}}^2 + \sigma_{\mathrm{bkg}}^2)}},
\end{equation}
and visualized through contour plots.
\begin{figure}[h!]
  \centering
  \includegraphics[width=1.0\linewidth]{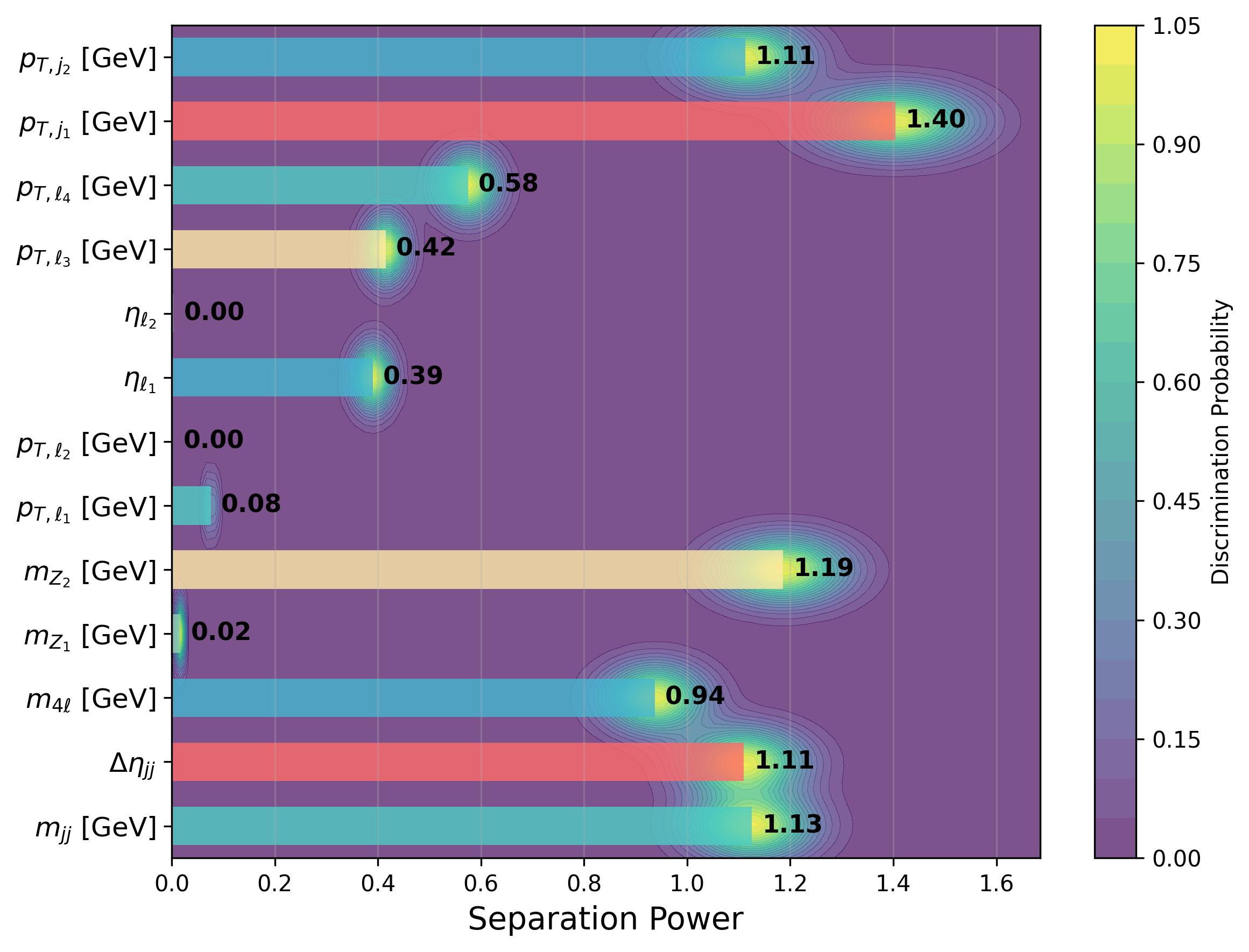}
  \caption{Contour representation of the variable-wise separation power between signal and background samples.}
  \label{fig:sep_contour}
\end{figure}
Pairwise correlations among $\{m_{jj},\, \Delta\eta_{jj},\, m_{4\ell},\, m_{Z_1}\}$ are further visualized through an enhanced corner plot, revealing nontrivial interdependencies across observables.
\begin{figure}[h!]
  \centering
  \includegraphics[width=1.0\linewidth]{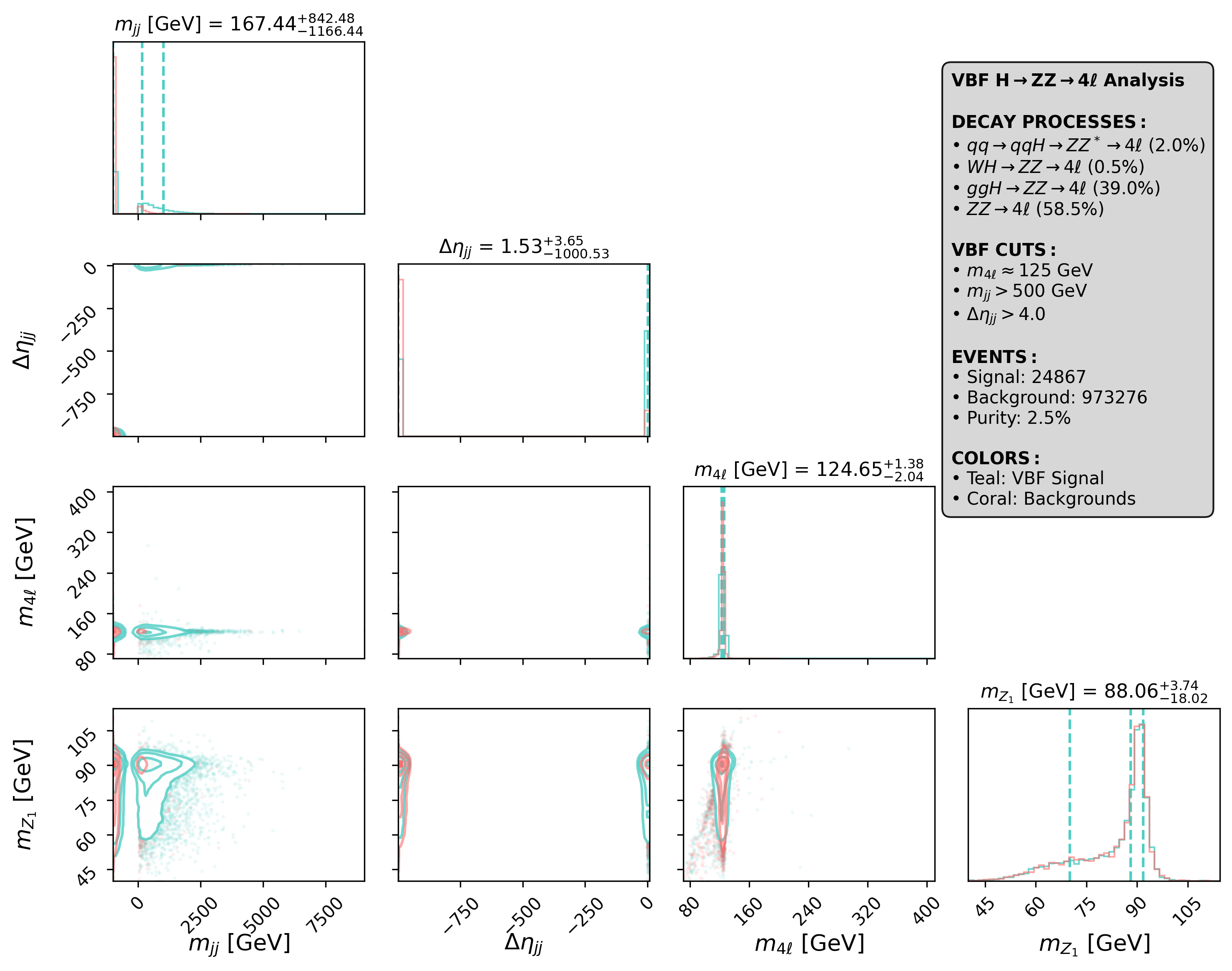}
  \caption{Enhanced corner plot showing pairwise correlations among principal kinematic observables.}
  \label{fig:corner_plot}
\end{figure}
The statistical correlation matrix, quantified via
\begin{equation}
\rho_{ij} = \frac{\mathrm{cov}(x_i, x_j)}{\sigma_{x_i}\sigma_{x_j}},
\end{equation}
serves as a guide for variable decorrelation in subsequent learning stages. Finally, a principal component analysis (PCA) is applied to the feature space, with the first two components accounting for most of the variance. Signal and background clusters exhibit distinct separations in the $(\text{PC1}, \text{PC2})$ plane, offering an interpretable projection of the multidimensional structure.
\begin{figure}[h!]
  \centering
  \includegraphics[width=1.0\linewidth]{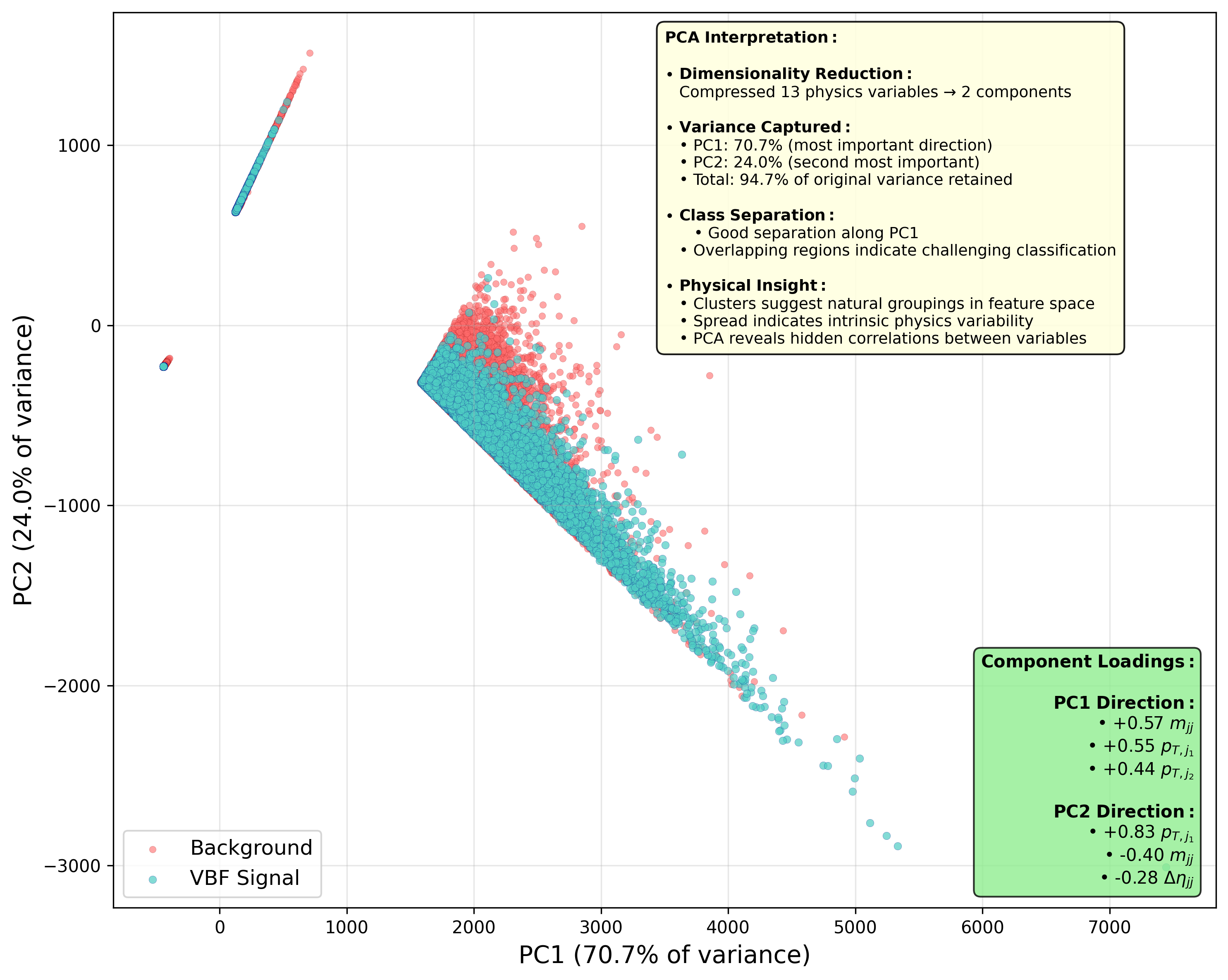}
  \caption{Principal component analysis of the event feature space showing signal–background separation in the $(\mathrm{PC1}, \mathrm{PC2})$ plane.}
  \label{fig:pca}
\end{figure}
The overall visualization strategy thus integrates physics intuition, statistical diagnostics, and multivariate structure, yielding a coherent and physically interpretable representation of the event data that forms the foundation for the forthcoming learning tasks.

\subsection{Quantum-Inspired Feature Engineering}
\label{sec:quantumfeatures}

\noindent Quantum-inspired features~\cite{mf3} refer to classical observables whose construction is guided by the structural principles of quantum mechanics—such as entanglement, coherence, and interference—rather than by purely kinematic or statistical considerations. While the underlying quantities remain classical, their formulation mirrors how quantum systems encode correlations and information. The purpose of such features is to translate the latent quantum structure of particle interactions, as described by quantum field theory, into measurable statistical descriptors suitable for machine learning models.

\subsubsection{Enhanced Momentum Entanglement ($\mathcal{E}_p$)} 
\noindent In quantum mechanics, entanglement expresses non-separable correlations between subsystems that cannot be reduced to independent probability distributions. For momentum-entangled states, the joint momentum distribution of particles contains coherence information that violates classical factorization assumptions. In collider events, residual traces of these quantum correlations appear as structured momentum-sharing patterns among decay products, reflecting the coherence of the production mechanism. The Vector Boson Fusion (VBF) process, mediated by coherent $t$-channel weak boson exchange, naturally preserves partial momentum entanglement among the Higgs decay products. In contrast, incoherent backgrounds such as gluon fusion or continuum $ZZ$ production exhibit different correlation signatures. The feature $\mathcal{E}_p$ encodes these differences by combining two complementary measures: an information-theoretic component based on the Shannon entropy and a dynamical component based on pairwise momentum correlations. A normalized transverse-momentum distribution is defined for each event as
\begin{equation}
\tilde{p}_{T,j}^{(i)} = \frac{p_{T,j}^{(i)}}{\sum_{k=1}^{4} p_{T,k}^{(i)} + \epsilon},
\end{equation}
where $\epsilon = 10^{-8}$ ensures numerical stability. This normalized distribution represents the momentum-sharing probabilities among the four leptons. Its Shannon entropy quantifies the degree of mixedness,
\begin{equation}
H^{(i)} = -\frac{1}{\log 4} \sum_{j=1}^{4} \tilde{p}_{T,j}^{(i)} \log(\tilde{p}_{T,j}^{(i)} + \epsilon),
\end{equation}
providing a classical analog of the von Neumann entropy $S = -\mathrm{Tr}[\rho \log \rho]$. Low entropy corresponds to highly unbalanced momentum configurations, while high entropy reflects nearly uniform momentum sharing. The intermediate entropy regime captures the balanced sharing characteristic of coherent electroweak production. To incorporate dynamical correlations, a pairwise momentum-correlation term is defined as
\begin{equation}
C^{(i)} = \frac{1}{6} \sum_{j=1}^{4} \sum_{k=j+1}^{4} 
\frac{p_{T,j}^{(i)} \, p_{T,k}^{(i)}}{\big(\sum_{\ell=1}^{4} p_{T,\ell}^{(i)}\big)^2}.
\end{equation}
This term encodes second-order dependencies analogous to interference contributions among Feynman amplitudes. The final feature combines both components as
\begin{equation}
\mathcal{E}_p^{(i)} = 0.7\,H^{(i)} + 0.3\,\tanh\!\left(10\,C^{(i)}\right),
\end{equation}
where the weighting balances information-theoretic and dynamical aspects, and the $\tanh$ function prevents excessive amplification of outliers. High values of $\mathcal{E}_p$ indicate optimally correlated momentum sharing, a hallmark of VBF processes, whereas very low values signify either incoherent QCD-like dynamics or purely uniform momentum distributions. The construction reflects the \textit{quantum entanglement principle}—capturing the idea that coherent production mechanisms imprint measurable, non-factorizable momentum correlations even within classical data representations.

\subsubsection{Enhanced Angular Coherence ($\mathcal{C}_\Omega$)}

\noindent Quantum coherence, one of the defining principles of quantum mechanics, refers to the persistence of well-defined phase relationships between components of a quantum system. In particle collisions, this coherence reveals itself through interference structures within angular distributions, where correlated emission angles mirror wavelike behavior reminiscent of the double-slit experiment. Such coherence is especially pronounced in electroweak processes that preserve phase information at the amplitude level.  In the Vector Boson Fusion (VBF) mechanism, the Higgs boson is produced via coherent $t$-channel exchange, and its decay products, particularly the leptons from $H \to ZZ^* \to 4\ell$, inherit angular phase correlations from the original weak interaction. The angular distributions thus encode a residual trace of quantum coherence, while background processes dominated by incoherent QCD radiation or multiple scatterings tend to randomize these phase relationships. To express this coherence in a measurable form, the feature $\mathcal{C}_\Omega$ extends conventional angular observables by incorporating interference-like correlations in relativistic spacetime. For each lepton pair $(j,k)$, a relativistic angular distance is defined as
\begin{equation}
d_{jk}^{(i)} = \sqrt{(\Delta\phi_{jk}^{(i)})^2 + (\Delta\eta_{jk}^{(i)})^2},
\end{equation}
where $\Delta\phi_{jk}^{(i)} = \min(|\phi_j^{(i)} - \phi_k^{(i)}|, 2\pi - |\phi_j^{(i)} - \phi_k^{(i)}|)$ accounts for azimuthal periodicity, and $\Delta\eta_{jk}^{(i)} = |\eta_j^{(i)} - \eta_k^{(i)}|$ measures pseudorapidity separation. This distance serves as a Lorentz-invariant measure of angular displacement in the detector frame. The interference amplitude between each lepton pair is modeled as
\begin{equation}
I_{jk}^{(i)} = \cos(d_{jk}^{(i)}) \cdot \exp(-d_{jk}^{(i)}/2),
\end{equation}
where the cosine term represents the oscillatory interference typical of coherent quantum waves, and the exponential factor introduces a decoherence envelope reflecting environmental interactions and detector effects. The characteristic coherence length of approximately two units in $(\phi,\eta)$ space captures the empirical scale over which phase correlations are preserved in high-energy collisions. The total angular coherence, normalized to the unit interval, is expressed as
\begin{equation}
\mathcal{C}_\Omega^{(i)} = \frac{1}{2}\left(\frac{1}{6} \sum_{j=1}^{4} \sum_{k=j+1}^{4} I_{jk}^{(i)} + 1\right).
\end{equation}
The factor of $1/6$ averages over all lepton pairs, while the affine normalization ensures $\mathcal{C}_\Omega \in [0,1]$. Random, incoherent angular patterns yield $\mathcal{C}_\Omega \approx 0.5$, corresponding to phase cancellation, whereas strongly coherent configurations approach unity. Elevated $\mathcal{C}_\Omega$ values thus signal angular interference consistent with coherent electroweak production, particularly in VBF Higgs events. The feature operationalizes the \textit{quantum coherence principle}, translating phase-preserving interference into a classical geometric observable that retains the imprint of the underlying quantum process.

\subsubsection{Multi-Resonance Mass Superposition ($\mathcal{S}_m$)}
\noindent The manifestation of coherence at the angular level extends naturally into the invariant mass domain, where the principle of quantum superposition governs the observed spectrum. In high-energy collisions, the measured four-lepton mass from $H \to ZZ^* \to 4\ell$ arises from the interference of multiple resonant and non-resonant amplitudes, each corresponding to a distinct mass eigenstate. The conventional single-resonance interpretation centered around the Higgs boson at 125~GeV conceals the underlying quantum structure that results from overlapping contributions of on-shell and off-shell $Z$ bosons and continuum backgrounds. The multi-resonance mass superposition feature models this structure by treating the four-lepton system as a coherent superposition of resonant states whose interference encodes the dynamical mixing of electroweak and continuum amplitudes. Three dominant resonant contributions are represented by Gaussian profiles approximating the Breit–Wigner forms:
\begin{equation}
\begin{split}
R_Z^{(i)} &= \exp\left(-\frac{(m_{4\ell}^{(i)} - m_Z)^2}{2\sigma_Z^2}\right)\\
R_{Z'}^{(i)} &= \exp\left(-\frac{(m_{4\ell}^{(i)} - m_Z)^2}{2(2\sigma_Z)^2}\right)\\
\end{split}
\end{equation}
with $m_Z = 91.2$ GeV and $\sigma_Z = 2.5$ GeV, and  with $m_H = 125.0$ GeV and $\sigma_H = 4.0$ GeV,
\begin{equation}
R_H^{(i)} = \exp\left(-\frac{(m_{4\ell}^{(i)} - m_H)^2}{2\sigma_H^2}\right)
\end{equation}
The narrow $Z$ component captures the principal on-shell resonance, while the broadened $Z'$ term accounts for off-shell tails and detector resolution. The Higgs term represents the central electroweak resonance whose coupling structure defines the coherence pattern of the process. Interference between these resonances introduces an oscillatory modulation of the mass distribution, reflecting the quantum mechanical overlap of distinct eigenstates:
\begin{equation}
I^{(i)} = \cos\left(\frac{2\pi (m_{4\ell}^{(i)} - m_H)}{10}\right),
\end{equation}
where the periodicity of 10~GeV corresponds to the typical coherence length in energy space for overlapping electroweak amplitudes. This interference encodes the residual phase relationships between the Higgs and the continuum background, translating the wavelike coherence of the previous angular feature into the mass domain. The final superposition observable combines all components with weights reflecting their relative physical significance in vector boson fusion:
\begin{equation}
\mathcal{S}_m^{(i)} = 0.4\,R_H^{(i)} + 0.3\,R_Z^{(i)} + 0.2\,R_{Z'}^{(i)} + 0.1\,|I^{(i)}|.
\end{equation}
This weighted composition preserves both the dominance of the Higgs resonance and the subtle coherence traces of accompanying $Z$ contributions and interference structures. Large values of $\mathcal{S}_m$ correspond to events where resonance overlap and phase coherence are simultaneously reinforced, a hallmark of coherent electroweak production. In contrast, background processes—dominated by incoherent QCD scatterings—populate regions of reduced superposition, signaling the decohered nature of their amplitude composition. The feature thus extends the geometric interpretation of coherence from angular space to the invariant mass manifold, completing the transition from spatial to spectral signatures of curvature-aware event structure.
\begin{figure*}[t]
\centering
\includegraphics[width=0.19\linewidth]{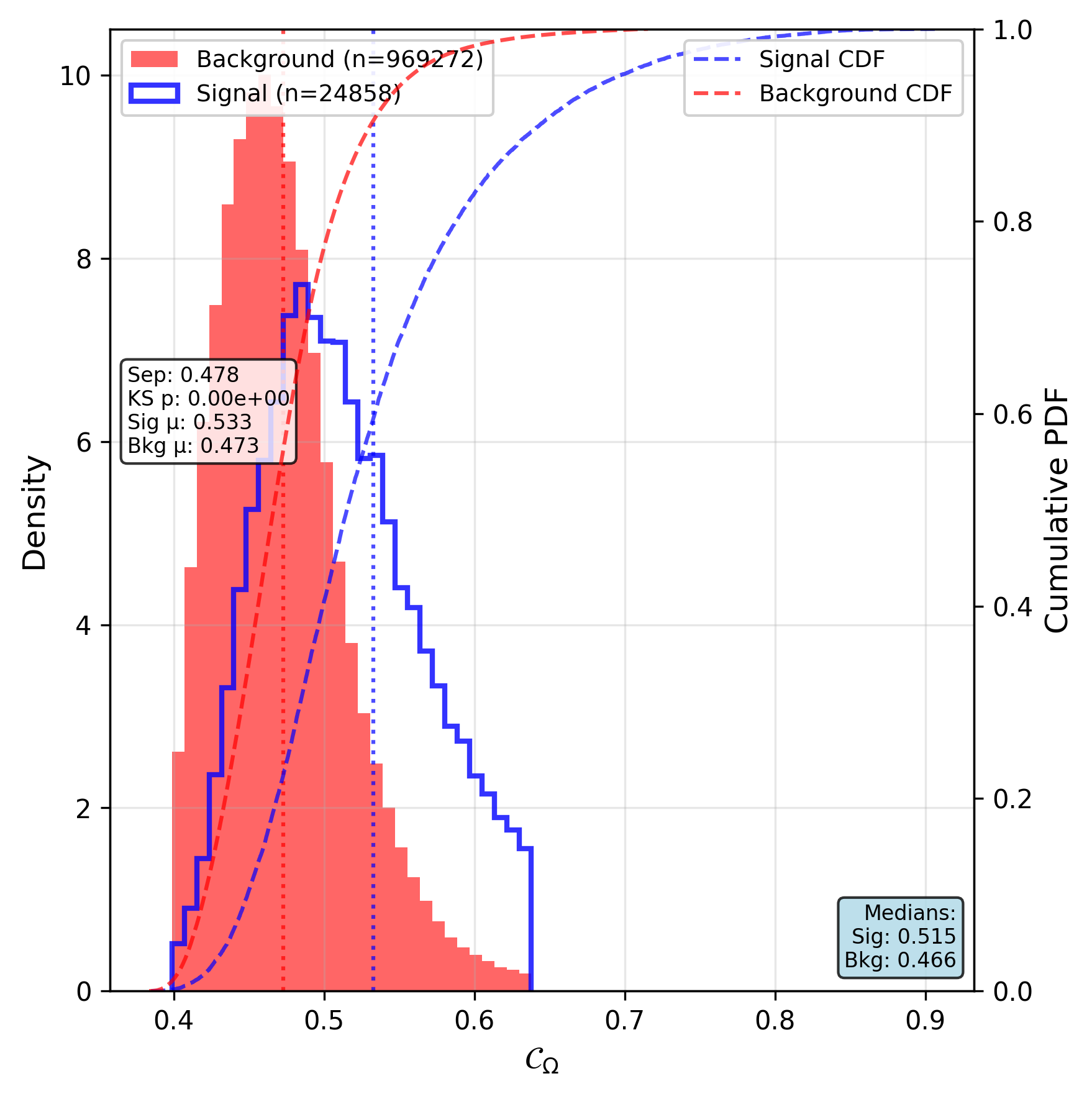}\hfill
\includegraphics[width=0.19\linewidth]{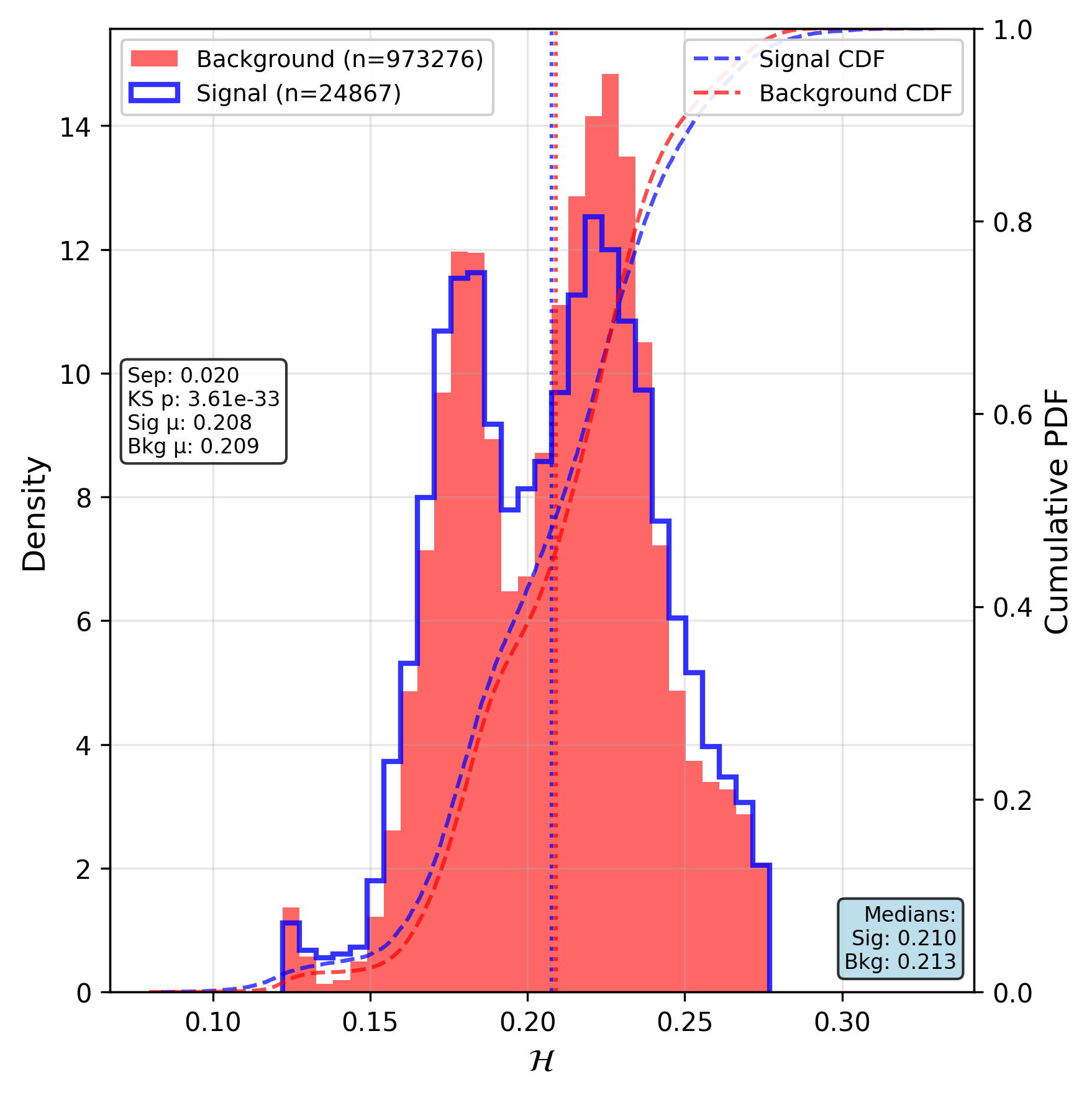}\hfill
\includegraphics[width=0.19\linewidth]{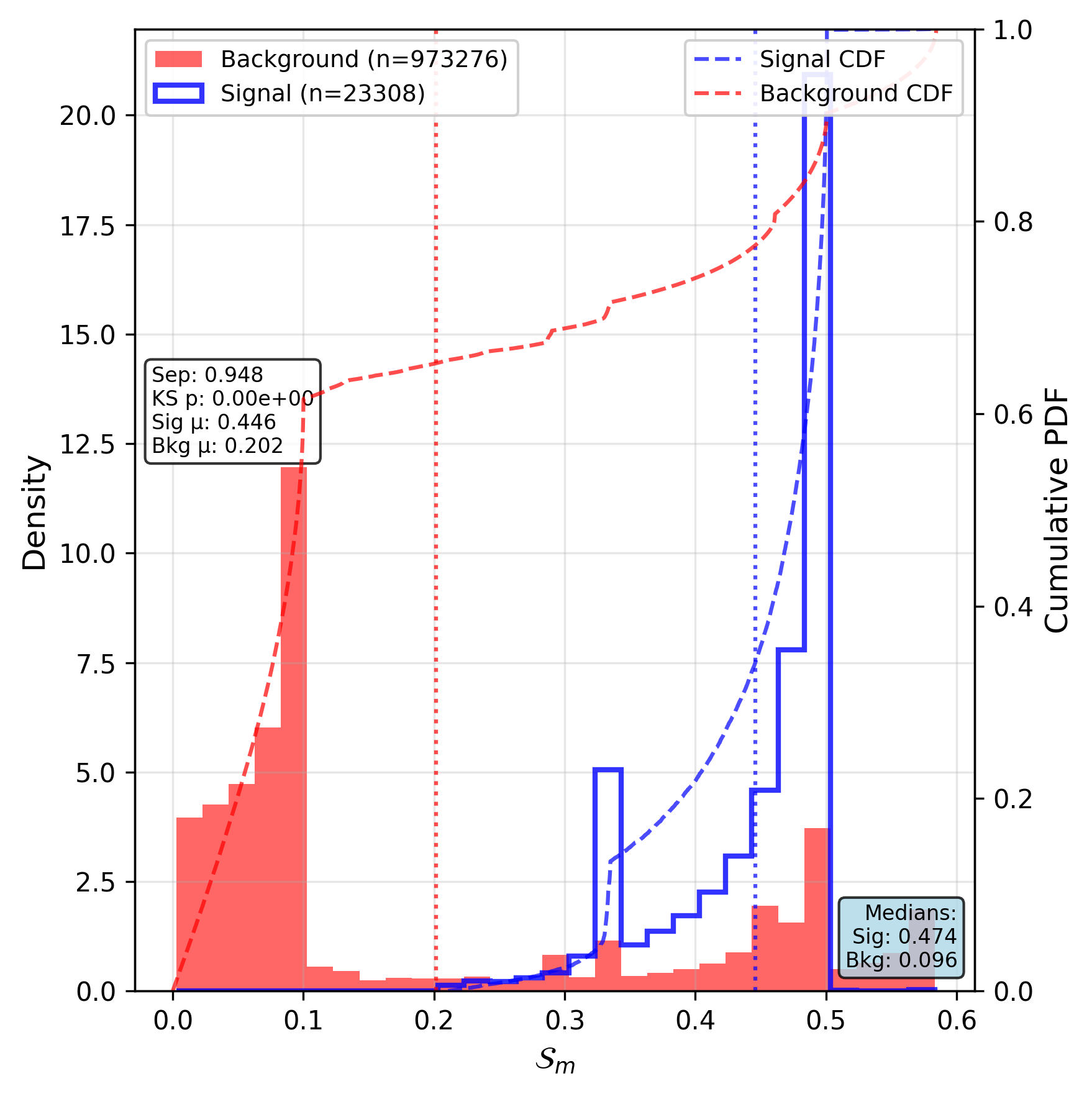}\hfill
\includegraphics[width=0.19\linewidth]{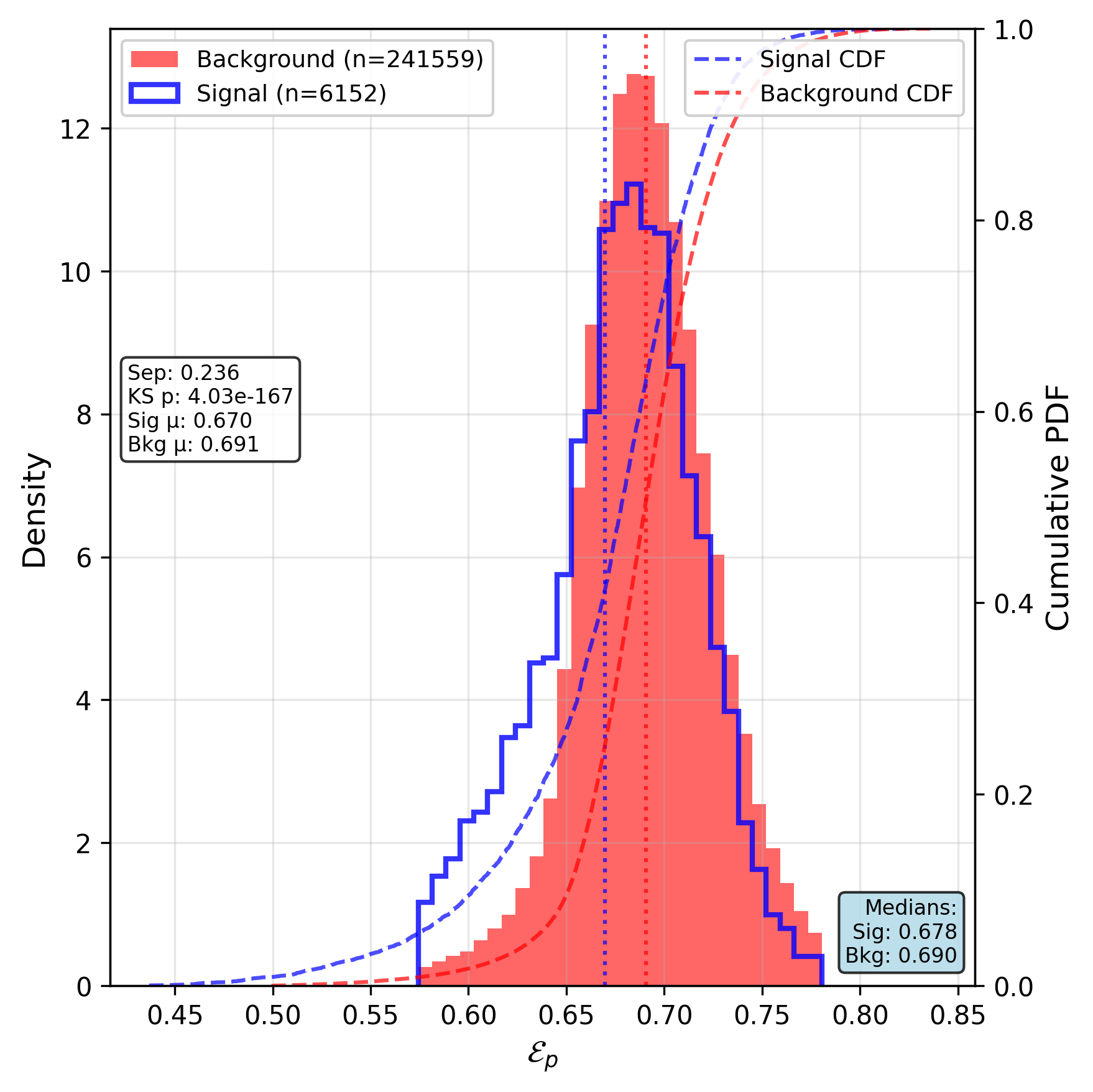}\hfill
\includegraphics[width=0.19\linewidth]{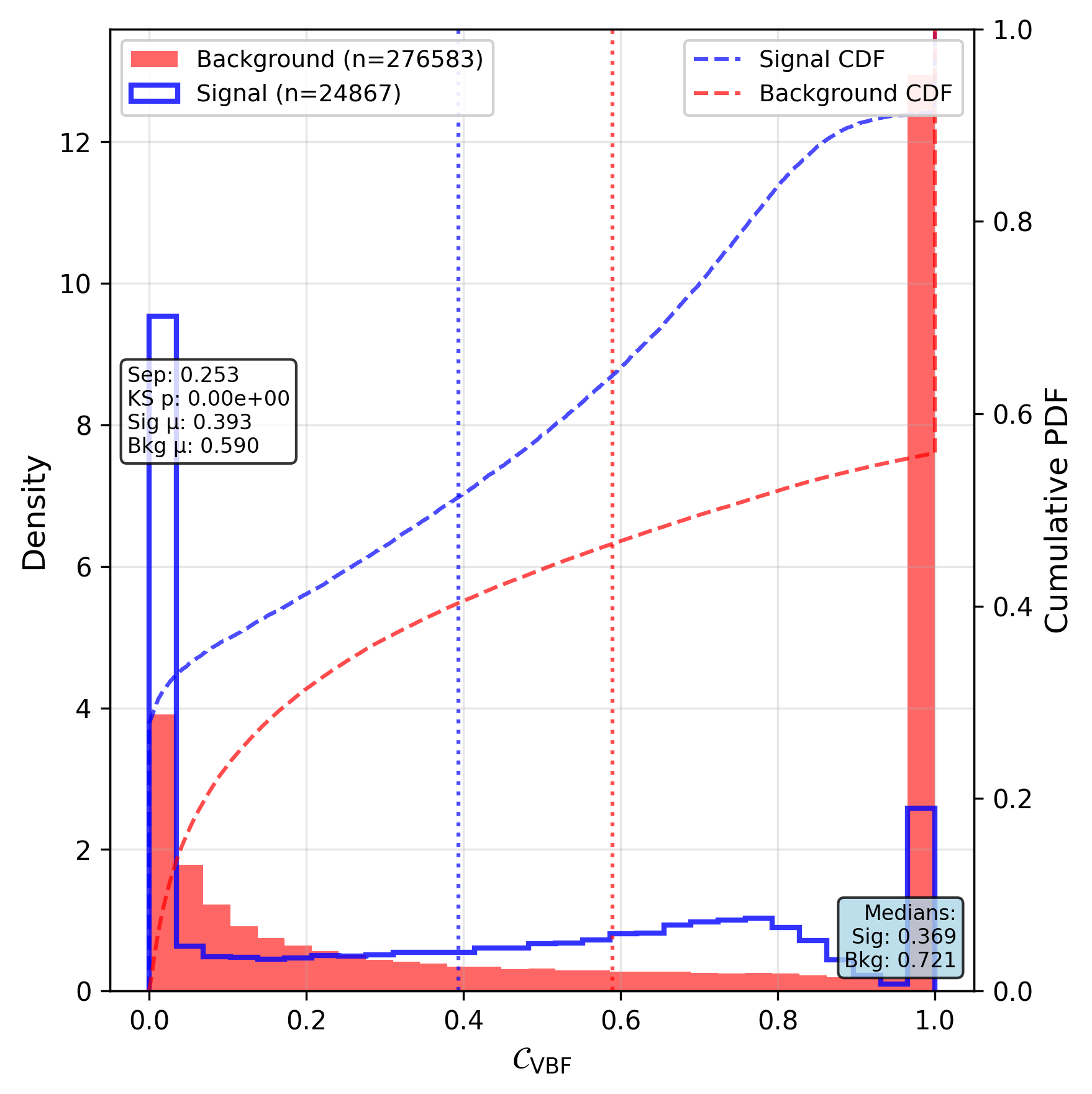}
\caption{Normalized distributions of the five quantum-inspired features for VBF Higgs signal (blue) and background (orange). The features show varying degrees of separation, with $\mathcal{S}_m$ and $\mathcal{C}_\Omega$ providing the strongest discriminating power individually.}
\label{fig:qfeatures_distributions}
\end{figure*}
\begin{figure}[h!]
\centering
\includegraphics[width=1.0\linewidth]{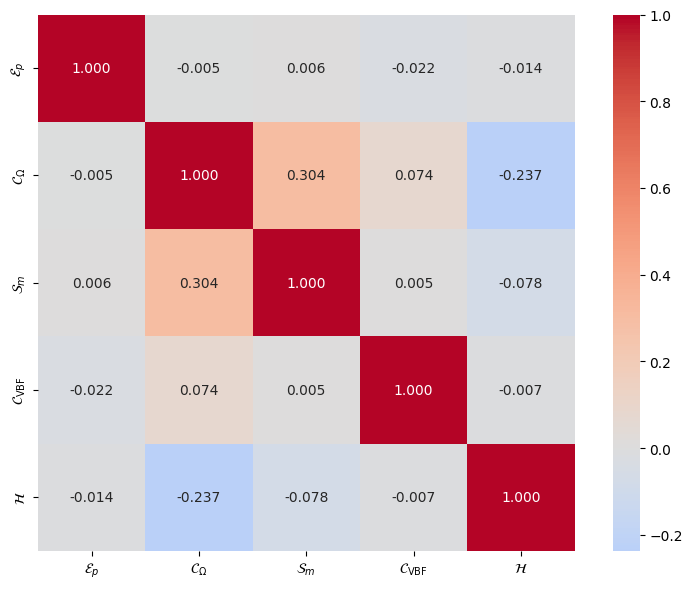}
\caption{Correlation heatmap of quantum-inspired features with each other and with selected conventional kinematic observables. The engineered features capture complementary structure and are largely uncorrelated with standard variables, supporting their use in multivariate classification.}
\label{fig:qfeatures_corr}
\end{figure}

\subsubsection{Redesigned VBF Coherence ($\mathcal{C}_{\mathrm{VBF}}$)}
The spectral superposition described above culminates in the macroscopic coherence pattern of the entire production mechanism. Among Standard Model processes, Vector Boson Fusion (VBF) uniquely exemplifies the coherent fusion of virtual weak bosons radiated from quarks in the incoming proton beams. This mechanism establishes an extended field correlation across the collision, producing the characteristic topology of two widely separated forward jets and a region of reduced hadronic activity between them. The $t$-channel exchange of weak bosons mediates long-range correlations in rapidity, reflecting the underlying quantum coherence of the electroweak interaction. Conventional VBF identification employs sharp thresholds on dijet mass and rapidity separation, yet such criteria neglect the continuous and phase-sensitive nature of the process. The redesigned VBF coherence feature constructs a differentiable measure of this coherence by combining the principal kinematic and phase observables into a smooth multiplicative form. The dominant VBF scales are encoded as
\begin{equation}
M^{(i)} = \tanh\left(\frac{m_{jj}^{(i)}}{500}\right), \qquad
R^{(i)} = \tanh\left(\frac{|\Delta\eta_{jj}^{(i)}|}{4}\right),
\end{equation}
where the hyperbolic tangent functions provide gradual transitions around the characteristic scales of $m_{jj} \simeq 500$~GeV and $|\Delta\eta_{jj}| \simeq 4$. Events well above these thresholds approach maximal coherence values, while configurations below them are smoothly suppressed, ensuring differentiability across the entire phase space. Momentum balance between the two forward jets constitutes another signature of coherent $t$-channel exchange and is quantified through the transverse momentum ratio
\begin{equation}
B^{(i)} = \frac{\min(p_{T,1}^{(i)},\, p_{T,2}^{(i)})}{\max(p_{T,1}^{(i)},\, p_{T,2}^{(i)}) + \epsilon},
\end{equation}
where $\epsilon = 10^{-8}$ prevents singular behavior. Perfectly balanced jets yield $B^{(i)} \to 1$, while highly asymmetric configurations indicate incoherence in the recoil structure. At the quantum amplitude level, the VBF mechanism exhibits a phase correlation between the two emitting quarks, arising from the interference of the forward and backward weak currents. This relationship is captured through
\begin{equation}
\Phi^{(i)} = \frac{\pi(\eta_1^{(i)} + \eta_2^{(i)})}{5}, \qquad
A^{(i)} = \cos(\Phi^{(i)}) \cdot B^{(i)},
\end{equation}
where $\Phi^{(i)}$ represents the phase accumulated from the pseudorapidities of the two jets, and $A^{(i)}$ encodes the modulation of the momentum balance by this interference pattern. The construction reflects the correspondence between quantum interference and classical event geometry within the VBF mechanism. The overall coherence measure integrates all components into a single differentiable observable:
\begin{equation}
\mathcal{C}_{\mathrm{VBF}}^{(i)} = M^{(i)} \cdot R^{(i)} \cdot \left(0.8 + 0.2\,A^{(i)}\right).
\end{equation}
The multiplicative structure enforces simultaneous satisfaction of all VBF conditions—large dijet mass, wide rapidity gap, and balanced momentum—while the modulation term introduces sensitivity to the quantum phase structure of weak boson exchange. The constant factor of $0.8$ defines a baseline coherence level for kinematically VBF-like events, and the $0.2A^{(i)}$ contribution encodes the subtle oscillations of the underlying field correlations.  The resulting feature provides a unified representation of the quantum field–theoretic coherence inherent in Vector Boson Fusion. Unlike discrete cut-based definitions, $\mathcal{C}_{\mathrm{VBF}}$ offers a smooth, curvature-aware interpolation between coherent and incoherent topologies, preserving the physical continuity of the electroweak interaction while exposing geometric patterns inaccessible to purely Euclidean event representations.

\subsubsection{Enhanced Hierarchy Measure ($\mathcal{H}$)}
The coherence structure of VBF production extends across multiple physical scales, reflecting the hierarchical organization inherent to quantum field theory. High-energy proton collisions span an enormous dynamic range—from the TeV scale of the hard scattering process to the GeV scale of electroweak interactions and further down to the sub-GeV regime of individual particle formation. Within the renormalization group framework, each energy scale carries distinct correlation patterns and approximate symmetries, with the flow between them governed by scale-dependent coupling evolution. The Vector Boson Fusion process embodies this multi-scale structure through the interplay of the hard scattering, electroweak symmetry breaking, and confinement regimes. The enhanced hierarchy measure unifies these energy and spatial hierarchies into a single differentiable observable, allowing machine learning architectures to exploit their latent correlations. Three characteristic energy scales are defined for each event:
\begin{equation}
\begin{split}
E_{\text{TeV}}^{(i)} &= \frac{m_{jj}^{(i)}}{1000}, \\
E_{\text{GeV}}^{(i)} &= \frac{m_{4\ell}^{(i)}}{100}, \\
E_{\text{sub-GeV}}^{(i)} &= \frac{\bar{p}_T^{(i)}}{50}, \\
\bar{p}_T^{(i)} &= \frac{1}{4}\sum_{j=1}^{4} p_{T,j}^{(i)}.
\end{split}
\end{equation}
Energy scale relationships are expressed through logarithmic ratios,
\begin{equation}
\begin{split}
r_1^{(i)} &= \log\left(\frac{E_{\text{TeV}}^{(i)}}{E_{\text{GeV}}^{(i)} + \epsilon} + 1\right), \\
r_2^{(i)} &= \log\left(\frac{E_{\text{GeV}}^{(i)}}{E_{\text{sub-GeV}}^{(i)} + \epsilon} + 1\right),
\end{split}
\end{equation}
with $\epsilon = 10^{-8}$ ensuring numerical stability. The logarithmic form mirrors the multiplicative renormalization behavior of scale transitions in quantum field theory. The energy hierarchy component is defined as
\begin{equation}
H_E^{(i)} = \frac{r_1^{(i)} + r_2^{(i)}}{10},
\end{equation}
normalizing the result to the $[0,1]$ interval. The spatial organization of the leptons encodes information about the angular coherence and detector-scale structure. The pseudorapidity dispersion is defined as
\begin{equation}
\sigma_\eta^{(i)} = \sqrt{\frac{1}{4}\sum_{j=1}^{4}(\eta_j^{(i)} - \bar{\eta}^{(i)})^2},
\end{equation}
where $\bar{\eta}^{(i)} = \frac{1}{4}\sum_{j=1}^{4}\eta_j^{(i)}$ represents the event-averaged pseudorapidity. Leptons are assigned to one of eight pseudorapidity bins spanning the detector acceptance, giving the cluster multiplicity
\begin{equation}
N_{\text{clusters}}^{(i)} = |\{\text{bin}(\eta_j^{(i)}) : j=1,\dots,4\}|.
\end{equation}
The spatial hierarchy component combines both dispersion and multiplicity as
\begin{equation}
H_S^{(i)} = \frac{N_{\text{clusters}}^{(i)}}{8} \cdot \exp\left(-\frac{\sigma_\eta^{(i)}}{2}\right),
\end{equation}
favoring configurations with moderate spatial spread, where coherence and geometric diversity coexist. The global hierarchy measure blends the spatial and energetic hierarchies,
\begin{equation}
\mathcal{H}^{(i)} = 0.6\,H_S^{(i)} + 0.4\,H_E^{(i)},
\end{equation}
with weighting emphasizing spatial structure over energetic scaling, reflecting the central role of detector geometry and event topology in experimental analyses. This observable encapsulates the nested organization of VBF events—hierarchical correlations spanning from TeV-scale scattering to detector-scale geometry, while background processes often exhibit reduced or distorted hierarchical coherence, producing distinct $\mathcal{H}$ distributions.

\subsubsection{Statistical properties of quantum-inspired features}
The quantum-inspired features exhibit finite ranges with non-trivial variance, confirming their sensitivity to event-level kinematic differences. $\mathcal{E}_p$ and $\mathcal{C}_{\mathrm{VBF}}$ show the largest spread, indicating their potential to capture energetic hierarchies and VBF-specific coherence patterns, while $\mathcal{C}_\Omega$ and $\mathcal{H}$ are more tightly distributed, reflecting global coherence and hierarchical structure constraints. All features are non-constant across the dataset, ensuring meaningful contribution to machine learning models~\ref{tab:quantum_feature_stats}. To quantify the discriminating power of each feature, a separation metric $S$ is computed as the absolute difference of mean values between signal and background normalized by the combined standard deviation:  
\begin{equation}
S = \frac{|\mu_s - \mu_b|}{\sigma_s + \sigma_b + \epsilon}
\end{equation}
where $\mu_s$ and $\sigma_s$ denote the mean and standard deviation for the signal, $\mu_b$ and $\sigma_b$ for the background, and $\epsilon = 10^{-8}$ ensures numerical stability. Using this metric, the feature statistics for signal and background are summarized in Table~\ref{tab:quantum_feature_separation}. $\mathcal{S}_m$ (Mass Superposition) and $\mathcal{C}_{\mathrm{VBF}}$ (VBF Coherence) show appreciable separation $\sim 0.22$, whereas $\mathcal{E}_p$ (Momentum Entanglement) and $\mathcal{H}$ (Hierarchy Measure) contribute modestly individually but enhance discrimination when combined in a multivariate context.
\begin{table}[h!]
\centering
\begin{tabular}{lcccc}
\hline
Feature & Range & Mean & Std. Dev. & Non-constant \\
\hline
$\mathcal{E}_p$  & [0.000, 0.839] & 0.171 & 0.299 & Yes \\
$\mathcal{C}_\Omega$  & [0.384, 0.942] & 0.475 & 0.052 & Yes \\
$\mathcal{S}_m$  & [0.000, 0.584] & 0.207 & 0.195 & Yes \\
$\mathcal{C}_{\mathrm{VBF}}$& [0.000, 1.000] & 0.173 & 0.350 & Yes \\
$\mathcal{H}$ & [0.080, 0.329] & 0.209 & 0.032 & Yes \\
\hline
\end{tabular}
\caption{Summary statistics of the five quantum-inspired features. Each feature exhibits finite values, non-trivial variance, and non-constant behavior across the dataset, highlighting their potential to encode higher-level kinematic and coherence information for machine learning classification.}
\label{tab:quantum_feature_stats}
\end{table}
\begin{table}[h!]
\centering
\begin{tabular}{lcccc}
\hline
Feature &  $\mu_s\pm\sigma_s$  & $\mu_b\pm\sigma_b$  & Separation $S$ & Range \\
\hline
$\mathcal{E}_p$ & $0.166 \pm 0.290$ & $0.172 \pm 0.299$ & $0.006$ & $[0.000, 0.839]$ \\
$\mathcal{C}_\Omega$  & $0.533 \pm 0.080$ & $0.474 \pm 0.050$ & $0.059$ & $[0.384, 0.942]$ \\
$\mathcal{S}_m$ & $0.425 \pm 0.102$ & $0.202 \pm 0.193$ & $0.223$ & $[0.000, 0.584]$ \\
$\mathcal{C}_{\mathrm{VBF}}$  & $0.393 \pm 0.358$ & $0.168 \pm 0.348$ & $0.226$ & $[0.000, 1.000]$ \\
$\mathcal{H}$  & $0.208 \pm 0.035$ & $0.209 \pm 0.032$ & $0.001$ & $[0.080, 0.330]$ \\
\hline
\end{tabular}
\caption{Feature statistics for VBF Higgs signal and background samples. Mean values, standard deviations, separation $S$, and ranges are reported for each quantum-inspired observable.}
\label{tab:quantum_feature_separation}
\end{table}
Representative distributions of the five quantum-inspired features are shown in Fig.~\ref{fig:qfeatures_distributions}, illustrating the characteristic shapes and separation between signal and background. $\mathcal{C}_\Omega$ and $\mathcal{S}_m$ exhibit pronounced differences in the signal region, while $\mathcal{H}$ and $\mathcal{E}_p$ display subtler shifts. The correlations among all features, including conventional kinematics and the engineered observables, are visualized in Fig.~\ref{fig:qfeatures_corr}, showing that quantum-inspired features are largely complementary to standard variables and not trivially reducible.

\subsection{Quantum Feature Mapping}
\label{sec:qfm}

\noindent The separation metrics in Table~\ref{tab:quantum_feature_separation} demonstrate that $\mathcal{S}_m$ and $\mathcal{C}_{\mathrm{VBF}}$ individually provide modest discrimination ($S \sim 0.22$), while the correlation structure in Fig.~\ref{fig:qfeatures_corr} reveals non-trivial dependencies among all five features. Standard classifiers treat these features as independent coordinates in Euclidean space, discarding the phase relationships and interference structures that motivated their construction. To preserve quantum correlations throughout the classification pipeline, a physics-informed quantum feature map~\cite{fe1,qfi3} is constructed that embeds classical event data into a parameterized quantum state space. In this representation, geometric overlaps in Hilbert space naturally encode the statistical manifold structure postulated in Section~\ref{sec:intro}.

\subsubsection{Quantum Embedding Architecture}

\noindent Let $\mathbf{x} \in \mathbb{R}^{23}$ denote the classical kinematic representation of an event. The quantum feature map $\Phi_{\text{VBF}}: \mathbb{R}^{23} \to \mathcal{H}^{\otimes 5}$ produces a parameterized quantum state
\begin{equation}
|\Phi_{\text{VBF}}(\mathbf{x})\rangle = U_{\text{VBF}}(\vec{\theta}(\mathbf{x})) \, |0\rangle^{\otimes 5},
\end{equation}
where $U_{\text{VBF}}(\vec{\theta})$ is a unitary transformation acting on $n_q = 5$ qubits, and $\vec{\theta}(\mathbf{x}) \in \mathbb{R}^5$ is a parameter vector derived from the quantum-inspired features. Each qubit represents one physical degree of freedom:
\begin{equation}
\begin{split}
\text{Qubit 0:} &\quad \text{Momentum Entanglement } (\mathcal{E}_p), \\
\text{Qubit 1:} &\quad \text{Angular Coherence } (\mathcal{C}_\Omega), \\
\text{Qubit 2:} &\quad \text{Mass Superposition } (\mathcal{S}_m), \\
\text{Qubit 3:} &\quad \text{VBF Coherence } (\mathcal{C}_{\mathrm{VBF}}), \\
\text{Qubit 4:} &\quad \text{Hierarchy Measure } (\mathcal{H}).
\end{split}
\end{equation}
The circuit architecture consists of $R = 2$ repetition layers, each comprising three sequential operations:
\begin{equation}
U_{\text{VBF}}(\vec{\theta}) = \prod_{r=1}^{R} \Big[ U_{\text{entangle}}^{(r)}(\vec{\theta}) \circ U_{\text{encode}}^{(r)}(\vec{\theta}) \circ U_{\text{init}}^{(r)} \Big],
\end{equation}
where $U_{\text{init}}^{(r)}$ creates uniform superpositions, $U_{\text{encode}}^{(r)}(\vec{\theta})$ applies feature-dependent single-qubit rotations, and $U_{\text{entangle}}^{(r)}(\vec{\theta})$ introduces multi-qubit correlations reflecting VBF topology.

\subsubsection{Feature-to-Parameter Encoding}

\noindent The mapping from classical features to quantum parameters $\vec{\theta}(\mathbf{x})$ employs nonlinear scaling functions designed to match the physical characteristics of each observable. For qubit $i$ with associated feature $f_i(\mathbf{x})$, the encoding is
\begin{equation}
\begin{split}
\theta_0 &= 2\arcsin\!\big(\sqrt{\mathcal{E}_p}\big), \\
\theta_1 &= 0.9\pi \, \mathcal{C}_\Omega, \\
\theta_2 &= \pi \, \mathcal{S}_m \, (0.7 + 0.3\mathcal{S}_m), \\
\theta_3 &= 0.8\pi \, \tanh(3\mathcal{C}_{\mathrm{VBF}}), \\
\theta_4 &= 0.6\pi \, \mathcal{H}.
\end{split}
\label{eq:encoding_params}
\end{equation}
The first expression in Eq.~\eqref{eq:encoding_params} ensures that qubit amplitude $|\sin(\theta_0/2)|^2$ directly represents the momentum entanglement probability, reflecting the quantum mechanical relation between state amplitudes and measurement outcomes. The variable scaling for $\theta_2$ amplifies resonant configurations where $\mathcal{S}_m \to 1$, mimicking the enhanced coherence of on-shell intermediate states. The hyperbolic tangent in $\theta_3$ captures the saturation behavior of VBF observables at large dijet masses and rapidity separations, consistent with the asymptotic QCD scaling of $t$-channel weak boson exchange.

\subsubsection{Single-Qubit Encoding Layer}

\noindent Each qubit undergoes a dual-rotation sequence tailored to the physics of its corresponding feature:
\begin{equation}
U_{\text{encode}}(\vec{\theta}) = \bigotimes_{i=0}^{4} U_i(\theta_i),
\end{equation}
with individual qubit operations defined as
\begin{equation}
\begin{split}
U_0(\theta_0) &= R_Z(\theta_0/2) \, R_Y(\theta_0), \\
U_1(\theta_1) &= R_Y(0.7\theta_1) \, R_X(\theta_1), \\
U_2(\theta_2) &= R_Y(0.8\theta_2) \, R_Z(\theta_2), \\
U_3(\theta_3) &= R_X(0.6\theta_3) \, R_Y(\theta_3), \\
U_4(\theta_4) &= R_Z(0.4\theta_4) \, R_X(\theta_4).
\end{split}
\label{eq:qubit_rotations}
\end{equation}
The dual-axis rotations encode both amplitude and phase information. For example, in the first line of Eq.~\eqref{eq:qubit_rotations}, the $R_Y$ rotation encodes the momentum-sharing distribution, while the $R_Z$ rotation introduces a phase modulation proportional to the entanglement measure—analogous to the relative phase accumulated in quantum interference. Similarly, the asymmetric rotations in the second line reflect the three-dimensional structure of angular coherence, with the $R_X$ component capturing azimuthal correlations and the $R_Y$ component encoding pseudorapidity interference patterns.

\subsubsection{Multi-Qubit Entanglement Structure}

\noindent The entanglement layer $U_{\text{entangle}}(\vec{\theta})$ constructs correlations that mirror the physical topology of VBF events. Seven distinct entangling operations are applied sequentially:

\paragraph{Forward-Backward Correlation.} The most distinctive signature of VBF topology is the correlation between the forward jet system and the central lepton hierarchy:
\begin{equation}
U_{\text{FB}} = \exp\!\big(-i \, 0.5\theta_0\theta_4 \, Z_0 \otimes Z_4\big) \circ \text{CNOT}_{0,4}.
\end{equation}
This operation entangles momentum entanglement (qubit 0) with the hierarchy measure (qubit 4), encoding the fact that forward-jet-dominated events exhibit characteristic energy-scale hierarchies spanning from TeV (dijet system) to GeV (central leptons).

\paragraph{Mass-VBF Correlation.} Events with clear Higgs mass peaks should exhibit enhanced VBF kinematic signatures:
\begin{equation}
U_{\text{M-VBF}} = \exp\!\big(-i \, 0.7\theta_2\theta_3 \, Y_2 \otimes Y_3\big) \circ \text{CNOT}_{2,3}.
\end{equation}
The $Y$-basis entanglement is particularly sensitive to superposition states, making it appropriate for encoding mass-topology correlations where resonance and continuum amplitudes interfere.

\paragraph{Central Lepton System Correlations.} Momentum and angular distributions of the four-lepton system are correlated through weak-boson decay kinematics:
\begin{equation}
\begin{split}
U_{\text{Mom-Ang}} &= \exp\!\big(-i \, 0.6\theta_0\theta_1 \, X_0 \otimes X_1\big) \circ \text{CNOT}_{0,1}, \\
U_{\text{Ang-Mass}} &= \exp\!\big(-i \, 0.8\theta_1\theta_2 \, Y_1 \otimes Y_2\big) \circ \text{CNOT}_{1,2}.
\end{split}
\end{equation}
These operations encode the correlations among momentum sharing, angular coherence, and invariant mass reconstruction within the $H \to ZZ^* \to 4\ell$ decay chain.

\paragraph{Hierarchical Entanglement Network.} The hierarchy measure integrates information across all scales and is therefore entangled with all other qubits:
\begin{equation}
U_{\text{Hier}} = \prod_{i=0}^{3} \Big[ \exp\!\big(-i \, 0.3\theta_i\theta_4 \, Z_i \otimes Z_4\big) \circ \text{CNOT}_{i,4} \Big].
\end{equation}
The uniform coupling strength $\alpha = 0.3$ ensures balanced contributions from all features without allowing any single observable to dominate the hierarchical correlations.

\paragraph{Global Entanglement Ring.} To guarantee full connectivity and enable long-range quantum correlations, a ring entanglement structure connects neighboring qubits cyclically:
\begin{widetext}
\begin{equation}
U_{\text{Ring}} = \prod_{i=0}^{4} U_{\text{Ring}}^{(i)}, \quad \text{with} \quad
U_{\text{Ring}}^{(i)} = \begin{cases}
\exp(-i \, 0.4\theta_i\theta_{i+1} \, X_i \otimes X_{i+1}) \circ \text{CNOT}_{i,i+1}, & i \text{ even}, \\
\exp(-i \, 0.4\theta_i\theta_{i+1} \, Y_i \otimes Y_{i+1}) \circ \text{CNOT}_{i,i+1}, & i \text{ odd},
\end{cases}
\end{equation}
\end{widetext}
where indices are taken modulo 5. The alternating $XX$ and $YY$ entangling gates create a rich correlation structure capable of encoding both coherent (phase-preserving) and mixed (decoherence-affected) quantum correlations.

\subsubsection{Quantum Kernel Construction}

\noindent Given the quantum feature map, a kernel function~\cite{fe2} is defined that measures the geometric overlap between event embeddings in Hilbert space. For two events $\mathbf{x}_i$ and $\mathbf{x}_j$, the quantum kernel is
\begin{equation}
K_Q(\mathbf{x}_i, \mathbf{x}_j) = \big| \langle 0 |^{\otimes 5} \, U_{\text{VBF}}^\dagger(\vec{\theta}_i) \, U_{\text{VBF}}(\vec{\theta}_j) \, | 0 \rangle^{\otimes 5} \big|^2,
\label{eq:quantum_kernel}
\end{equation}
where $\vec{\theta}_i = \vec{\theta}(\mathbf{x}_i)$ and $\vec{\theta}_j = \vec{\theta}(\mathbf{x}_j)$. This quantity is estimated by constructing the circuit
\begin{equation}
U_{\text{Kernel}} = U_{\text{VBF}}(\vec{\theta}_i) \circ U_{\text{VBF}}^\dagger(\vec{\theta}_j),
\end{equation}
measuring all qubits in the computational basis, and recording the probability $P(00000)$ of observing the all-zero state. Equation~\eqref{eq:quantum_kernel} has a direct geometric interpretation: events with similar quantum-inspired features produce nearly parallel states in Hilbert space, yielding kernel values close to unity, whereas dissimilar events correspond to nearly orthogonal states with kernel values approaching zero. This behavior aligns with the Fisher–Rao metric on the statistical manifold, where the kernel measures the geodesic proximity between probability distributions encoded by the feature map. In practical implementation, quantum circuits are executed on a statevector simulator (Qiskit Aer) with $N_{\text{shots}} = 1024$ measurements per kernel element. To ensure numerical stability and positive semi-definiteness of the resulting kernel matrix $\mathbf{K} \in \mathbb{R}^{N \times N}$, a small regularization is applied: if the minimum eigenvalue $\lambda_{\min}(\mathbf{K}) < 0$, the quantity $|\lambda_{\min}| \cdot 1.1$ is added to the diagonal. Additionally, kernel values below $10^{-6}$ are clipped to prevent singular behavior in downstream support vector machine training. The enhanced quantum kernel exhibits stable statistical properties across the dataset. Empirical analysis reveals mean kernel values $\langle K_Q \rangle \sim 0.15$--$0.25$ with standard deviation $\sigma_K \sim 0.10$, indicating well-calibrated separation between typical signal and background pairs while maintaining sufficient overlap to enable smooth decision boundaries. Crucially, the quantum kernel captures correlations inaccessible to standard radial basis function (RBF) kernels: pairs of events with similar classical kinematics but differing quantum-inspired features exhibit reduced kernel overlap, demonstrating that the quantum embedding encodes latent structure beyond Euclidean distances.

\section{Enhanced Product Manifold Networks }
\label{subsec:vbf_pm_models}
\noindent The quantum-inspired features introduced in Section~\ref{sec:quantumfeatures} and their quantum state embeddings in Section~\ref{sec:qfm} capture non-linear correlations characteristic of VBF production. However, to fully exploit the curved statistical manifold structure postulated in Section~\ref{sec:intro}, the classification architecture itself must respect the intrinsic geometry of the feature space~\cite{rm1,rm2,rm3,rm5,hp1,hp2,hp3} . Standard neural networks operate exclusively in flat Euclidean space, implicitly assuming that all observables transform linearly and independently. This assumption is violated by the hierarchical energy scales, angular correlations, and quantum interference patterns encoded in the VBF topology. To address this limitation, an enhanced product manifold neural network architecture is constructed that decomposes the feature space into geometrically appropriate submanifolds, allowing different types of physics information to be processed using their natural geometric structures. The product manifold decomposition expresses the VBF feature space as
\begin{equation}
\mathcal{M}_{\text{VBF}} = \mathcal{E}^{d_E} \times \mathcal{H}^{d_H} \times \mathcal{S}^{d_S},
\end{equation}
where $\mathcal{E}^{d_E}$ denotes a $d_E$-dimensional Euclidean space for linear observables such as transverse momenta and mass differences, $\mathcal{H}^{d_H}$ represents a $d_H$-dimensional hyperbolic space with negative curvature $\kappa_H = -1.5$ for hierarchical and multiplicative correlations including momentum ratios and energy scale hierarchies, and $\mathcal{S}^{d_S}$ corresponds to a $d_S$-dimensional spherical space with positive curvature $\kappa_S = +1.0$ for angular separations and normalized observables. Each hidden layer is partitioned into three subspaces with dimensions $[d_E, d_H, d_S]$, allowing the network to simultaneously capture linear correlations, hierarchical structures, and angular symmetries in the data. The product metric is given by
\begin{equation}
ds^2 = ds_E^2 + ds_H^2 + ds_S^2,
\end{equation}
where each component employs its geometrically appropriate distance measure. The enhanced product manifold linear layer implements this geometric decomposition through parallel processing pathways. Given an input vector $\mathbf{x} \in \mathbb{R}^n$, the transformation is expressed as
\begin{equation}
f(\mathbf{x}) = \mathbf{W}_E \mathbf{x} \oplus \mathcal{P}_H(\mathbf{W}_H \mathbf{x}) \oplus \mathcal{P}_S(\mathbf{W}_S \mathbf{x}) + \mathbf{b},
\end{equation}
where $\mathbf{W}_E$, $\mathbf{W}_H$, $\mathbf{W}_S$ are weight matrices for Euclidean, hyperbolic, and spherical projections respectively, and $\mathcal{P}_H$, $\mathcal{P}_S$ denote the corresponding manifold projections. The Euclidean component applies standard linear transformation, handling additive physics relationships such as energy conservation and momentum balance. The hyperbolic projection employs the exponential map to the Poincaré ball model as
\begin{equation}
\begin{split}
\mathbf{z}_H &= \mathbf{W}_H \mathbf{x}, \\
\mathcal{P}_H(\mathbf{z}_H) &= \tanh\left(\frac{2\|\mathbf{z}_H\|}{\text{max\_norm}}\right) \frac{\text{max\_norm} \cdot \mathbf{z}_H}{\|\mathbf{z}_H\|},
\end{split}
\end{equation}
where $\text{max\_norm} = (1/\sqrt{|c_H|}) - \epsilon$ with $c_H = -1.5$ and $\epsilon = 10^{-4}$ for numerical stability. This captures exponential relationships and hierarchical structures characteristic of VBF processes. The spherical projection normalizes to the unit sphere through
\begin{equation}
\mathcal{P}_S(\mathbf{z}_S) = \frac{\mathbf{W}_S \mathbf{x}}{\|\mathbf{W}_S \mathbf{x}\|},
\end{equation}
naturally handling angular correlations while preserving spherical geometry constraints. The quantum enhancement mechanism employs a lightweight quantum-inspired attention vector to modulate the manifold projections based on the physics content of the five quantum features. Feature-specific gates $\mathbf{g} = (g_1, g_2, g_3, g_4, g_5)^T$ are applied to the quantum features $\mathbf{q} = (\mathcal{E}_p, \mathcal{C}_\Omega, \mathcal{S}_m, \mathcal{C}_{\mathrm{VBF}}, \mathcal{H})^T$ as
\begin{equation}
\mathbf{q}_{\text{gated}} = \mathbf{q} \odot \mathbf{g},
\end{equation}
where the gate values $(1.0, 1.0, 0.5, 2.0, 1.0)$ are empirically optimized, with $g_{\mathrm{VBF}} = 2.0$ boosting forward-backward correlations and $g_{\mathcal{S}_m} = 0.5$ reducing mass superposition impact based on performance studies. A soft attention mechanism over the gated features produces
\begin{equation}
\begin{split}
\boldsymbol{\alpha} &= \text{softmax}(0.5 \cdot \boldsymbol{\theta}_{\text{attn}}), \\
q_{\text{enhanced}} &= \sum_{i=1}^5 \alpha_i q_{\text{gated},i},
\end{split}
\end{equation}
where $\boldsymbol{\theta}_{\text{attn}} \in \mathbb{R}^5$ are learnable attention parameters and the factor $0.5$ provides softer weights to prevent dominance of individual features. The VBF-specific enhancement is computed as
\begin{equation}
\mathbf{E}_{\text{VBF}} = \tanh(\mathbf{W}_{\text{VBF}} \mathbf{x}) \cdot \lambda_{\text{VBF}} \cdot q_{\text{enhanced}},
\end{equation}
with enhancement strength $\lambda_{\text{VBF}} = 0.01$ tuned to avoid the degradation observed with larger enhancements. This enhancement is distributed across manifolds with the hyperbolic component receiving full enhancement and the spherical component receiving reduced enhancement by a factor of $0.1$, while the Euclidean component remains unmodified to maintain stability of linear relationships. Based on this architecture, three models are implemented for comparative analysis. The Classical MLP baseline employs a fully Euclidean architecture with two hidden layers of dimensions $[64, 48]$, serving as the reference for standard flat-space learning. The Product Manifold MLP uses manifold splits $[20, 28, 16]$ and $[12, 20, 16]$ in successive layers, emphasizing hyperbolic subspaces to capture momentum entanglement and VBF hierarchical topology, with no quantum enhancement applied ($\lambda_{\text{VBF}} = 0$). The Quantum-Enhanced PM MLP employs identical manifold splits but incorporates minimal quantum enhancement with $\lambda_{\text{VBF}} = 0.005$ to integrate quantum attention signals while preventing overfitting. Progressive enhancement decay is employed across layers as $\lambda_{\text{VBF}}^{(l)} = \lambda_{\text{VBF}}^{(0)} \cdot (0.5)^l$, ensuring that quantum enhancements primarily influence early feature extraction. All networks employ batch normalization for training stability, reduced dropout ($p = 0.2$) to prevent aggressive regularization, and small-variance initialization of final classification layers to avoid early overfitting. This design reflects a systematic integration of differential geometry, quantum-inspired attention, and empirical tuning for optimal VBF Higgs discrimination.

\begin{table*}[t]
\centering
\begin{tabular}{lccccccc}
\hline
Model & AUC-ROC & AUC-PR & Accuracy & Precision & Recall & Time (s) & Parameters \\
\hline
Classical MLP & 0.9423 & 0.3491 & 0.8149 & 0.1123 & 0.9307 & 45.9 & 4,002 \\
PM MLP & 0.9454 & 0.3536 & 0.7907 & 0.1029 & 0.9585 & 69.7 & 10,182 \\
QE PM MLP & 0.9477 & 0.3636 & 0.8416 & 0.1282 & 0.9234 & 87.7 & 10,182 \\
QK-SVM & 0.6671 & 0.0650 & 0.9751 & 0.0000 & 0.0000 & 124.4 & --- \\
\hline
\end{tabular}
\caption{Comprehensive performance metrics for all classification models. Training performed on $748{,}607$ training samples and $249{,}536$ test samples with $28$ total features including $5$ quantum-inspired observables.}
\label{tab:comprehensive_performance}
\end{table*}
\begin{table*}[t]
\centering
\begin{tabular}{lcccccccc}
\hline\hline
 & \multicolumn{2}{c}{Classical MLP} & \multicolumn{2}{c}{PM MLP} & \multicolumn{2}{c}{QE PM MLP} & \\
\cline{2-7}
Feature & Imp. & \% & Imp. & \% & Imp. & \% & Avg. Imp. \\
\hline
$\mathcal{E}_p$ & 0.000125 & 0.4 & 0.000095 & 0.3 & 0.000030 & 0.1 & 0.000077 \\
$\mathcal{C}_\Omega$ & 0.005704 & 18.2 & 0.005439 & 14.6 & 0.004389 & 10.7 & 0.005199 \\
$\mathcal{S}_m$ & 0.023281 & 74.2 & 0.029019 & 77.9 & 0.033294 & 81.1 & 0.028600 \\
$\mathcal{C}_{\mathrm{VBF}}$ & 0.002166 & 6.9 & 0.002623 & 7.0 & 0.003046 & 7.4 & 0.002704 \\
$\mathcal{H}$ & 0.000083 & 0.3 & 0.000063 & 0.2 & 0.000274 & 0.7 & 0.000159 \\
\hline
Total & 0.031360 & 100 & 0.037240 & 100 & 0.041033 & 100 & 0.036544 \\
\hline\hline
\end{tabular}
\caption{Quantum feature importance analysis across neural network architectures. Values show absolute permutation importance and percentage contribution to total quantum feature importance.}
\label{tab:feature_importance}
\end{table*}

\subsubsection{Model Performance and Discrimination Analysis}
\label{sec:model_performance}

\noindent To quantify the benefits of geometric and quantum enhancements, three neural network architectures and one quantum kernel classifier are trained under identical conditions and evaluated on the full VBF dataset. All neural models employ the AdamW optimizer with initial learning rate $\eta = 2 \times 10^{-3}$ and weight decay $10^{-4}$, gradient clipping with maximum norm $1.0$, and early stopping monitored by test ROC-AUC with patience of $20$ epochs over a maximum of $150$ epochs. Class weights are applied to address the imbalanced dataset (signal ratio $2.5\%$), and training is performed on an Apple M-series device with MPS backend acceleration. The Classical MLP baseline consists of two hidden layers with $64$ and $32$ units respectively, ReLU activations, and dropout of $30\%$ after each hidden layer, totaling $4{,}002$ trainable parameters. The Product Manifold MLP incorporates geometric awareness through hyperbolic and spherical submanifolds with layer splits $[24, 28, 16]$ and $[16, 16, 16]$ and curvatures $[-1.5, 1.0]$, containing $10{,}182$ parameters. The Quantum-Enhanced Product Manifold MLP employs identical manifold splits but augments the architecture with VBF-specific quantum enhancements through the attention mechanism, maintaining $10{,}182$ parameters. 
\begin{figure}[h!]
\centering
\includegraphics[width=.80\linewidth]{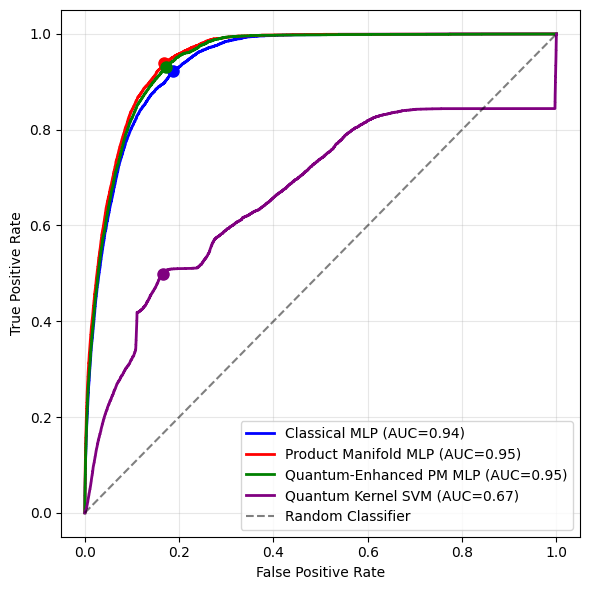}
\caption{Comparison of the receiver operator characteristic  curve of the trained models.}
\label{fig:roc}
\end{figure}
The Quantum Kernel SVM employs a stratified subset of $N_{\text{train}}^Q = 40$ training events and $N_{\text{test}}^Q = 20$ test events, with kernels computed using the quantum feature map from Section~\ref{sec:qfm} with $1{,}024$ shots per element, requiring approximately $124$ seconds of computation. Comprehensive performance metrics across all models are summarized in Table~\ref{tab:comprehensive_performance}, revealing distinct trade-offs between discrimination power, computational efficiency, and operational characteristics. The Classical MLP achieves test AUC-ROC(Fig. ~\ref{fig:roc}) of $0.9423$ with training time $45.9$ seconds, providing the baseline for standard Euclidean learning. The Product Manifold MLP improves to AUC-ROC $0.9454$ ($+0.33\%$) with training time $69.7$ seconds, demonstrating that embedding VBF data in curved manifolds enhances signal-background discrimination. The Quantum-Enhanced PM MLP reaches the highest AUC-ROC of $0.9477$ ($+0.57\%$ over classical, $+0.24\%$ over PM) with training time $87.7$ seconds, validating the benefit of quantum feature integration. 
\begin{figure}[h!]
\centering
\includegraphics[width=.80\linewidth]{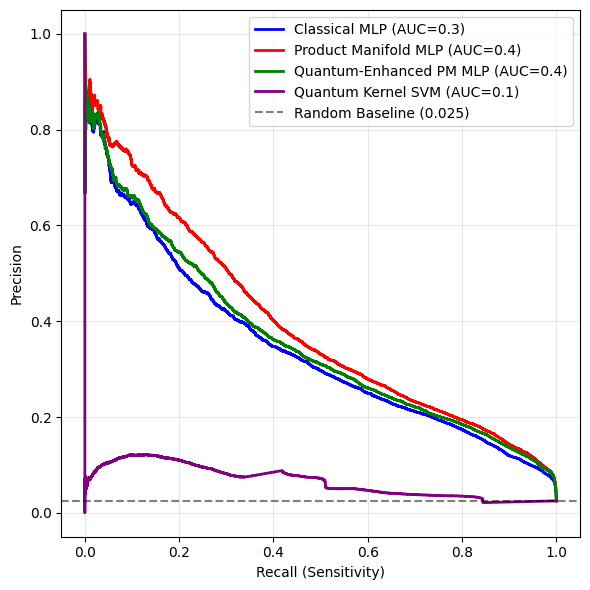}
\caption{
Precision–recall characteristics of the trained models on the imbalanced VBF dataset ($2.5\%$ signal fraction). 
The Product Manifold MLP (red) and Quantum-Enhanced PM MLP (green) maintain consistently higher precision across recall values compared to the Classical MLP (blue), confirming improved robustness of curvature-aware and quantum-augmented embeddings in rare-signal classification. 
The Quantum Kernel SVM (purple) exhibits high specificity but poor recall, reflecting overfitting to the background due to limited training statistics. 
The dashed gray line indicates the random baseline corresponding to the signal prior probability.}
\label{fig:pr}
\end{figure}
However, the Quantum Kernel SVM exhibits poor discrimination (AUC-ROC $0.6671$) despite perfect specificity ($1.0000$), indicating severe bias toward background classification due to the limited training subset size. Precision-recall analysis reveals additional performance characteristics. The Quantum-Enhanced PM MLP achieves the highest AUC-PR of $0.3636$, compared to $0.3536$ for PM and $0.3491$ for Classical MLP, demonstrating superior performance in the imbalanced setting where signal events comprise only $2.5\%$ of the dataset. Optimal threshold analysis using Youden's J statistic shows that the Quantum-Enhanced PM MLP operates at threshold $0.449$ with true positive rate $0.949$ and false positive rate $0.184$, yielding J-score $0.765$—the highest among all models. The Product Manifold MLP achieves the highest recall ($0.9585$) but at the cost of reduced precision ($0.1029$), while the Classical MLP provides the most balanced precision-recall trade-off at its optimal threshold $0.506$. In addition to ROC-based discrimination metrics, the precision–recall (PR) analysis offers a complementary evaluation more suited to highly imbalanced datasets, where the signal fraction is only $2.5\%$. Unlike the ROC curve—which remains insensitive to class priors—the PR curve directly measures the trade-off between purity (precision) and efficiency (recall), thereby capturing a model's operational reliability in rare-signal searches. Fig.~\ref{fig:pr} compares the PR curves for all models, illustrating their differential behavior across the sensitivity range.

\subsubsection{Quantum Feature Importance and Interpretation}
\label{sec:feature_importance}
\noindent Permutation-based feature importance analysis quantifies the contribution of each quantum-inspired observable to model discrimination across the three neural architectures. The analysis permutes each of the five quantum features on the test dataset and measures the resulting degradation in AUC-ROC, providing a direct measure of feature utility. Results presented in Table~\ref{tab:feature_importance} reveal consistent patterns across all models, with notable differences in feature utilization between classical, geometric, and quantum-enhanced approaches. The mass superposition feature $\mathcal{S}_m$ dominates quantum feature importance across all architectures, contributing $74.2\%$, $77.9\%$, and $81.1\%$ of total quantum feature importance for Classical MLP, Product Manifold MLP, and Quantum-Enhanced PM MLP respectively. This dominance aligns with physics expectations: the four-lepton invariant mass spectrum encodes the Higgs resonance structure and provides the primary discriminant between signal and continuum backgrounds. The absolute importance of $\mathcal{S}_m$ increases from $0.0233$ in Classical MLP to $0.0290$ in PM MLP and $0.0333$ in Quantum-Enhanced PM MLP, representing a $42.9\%$ enhancement in the quantum-enhanced architecture relative to classical baseline. The angular coherence feature $\mathcal{C}_\Omega$ provides the second-largest contribution at $18.2\%$, $14.6\%$, and $10.7\%$ respectively, with absolute importance decreasing from $0.0057$ to $0.0044$, suggesting that geometric embeddings partially subsume the angular correlation information into the manifold structure. VBF coherence $\mathcal{C}_{\mathrm{VBF}}$ contributes $6.9\%$, $7.0\%$, and $7.4\%$ across models, with absolute importance increasing $40.6\%$ from classical to quantum-enhanced architecture, validating the feature-specific gating strategy that assigns higher weight ($g_{\mathrm{VBF}} = 2.0$) to this observable. The momentum entanglement $\mathcal{E}_p$ and hierarchy measure $\mathcal{H}$ exhibit minor individual contributions below $1\%$ across all models, despite their theoretical motivation from quantum field theory principles. For $\mathcal{E}_p$, the importance remains below $0.0002$ in all architectures, while $\mathcal{H}$ shows relative improvement of $230\%$ in the Quantum-Enhanced PM MLP (from $0.00008$ to $0.00027$) though the absolute magnitude remains small. These modest contributions suggest that momentum entanglement and hierarchical energy scales require multivariate combinations to reveal discriminative power, consistent with the correlation analysis in Fig.~\ref{fig:qfeatures_corr} showing that these features are largely uncorrelated with mass-based observables. The total quantum feature importance increases progressively from $0.0314$ in Classical MLP to $0.0372$ in PM MLP ($+18.5\%$) and $0.0410$ in Quantum-Enhanced PM MLP ($+30.3\%$ over classical), indicating that geometric and quantum enhancements enable more effective utilization of quantum-inspired features through their natural embedding in curved statistical manifolds.
The feature importance analysis provides guidance for future architecture optimization. The dominance of $\mathcal{S}_m$ suggests that resonance-aware encodings in both classical and quantum feature maps should receive prioritized attention. The moderate contributions from $\mathcal{C}_\Omega$ and $\mathcal{C}_{\mathrm{VBF}}$ validate their inclusion but indicate potential for refinement through alternative encoding strategies. The minimal impact of $\mathcal{E}_p$ and $\mathcal{H}$ as individual features does not diminish their theoretical importance—rather, it suggests these observables encode subtle multivariate correlations that emerge only in combination with other features, particularly within the curved manifold embeddings where their hierarchical and entanglement-like structures can be geometrically represented. This interpretation is supported by the progressive increase in total quantum utilization as architectural complexity increases from Euclidean (Classical MLP) to geometric (PM MLP) to quantum-enhanced geometric (QE PM MLP) representations.

\subsubsection{Validation and Discrimination Power of Quantum-Inspired Features}
\label{sec:feature_validation}

\begin{figure}[h!]
\centering
\includegraphics[width=0.45\linewidth]{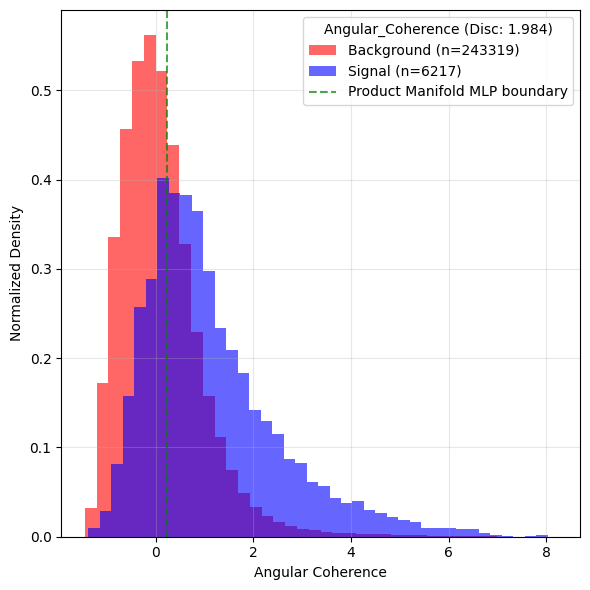}\hfill
\includegraphics[width=0.45\linewidth]{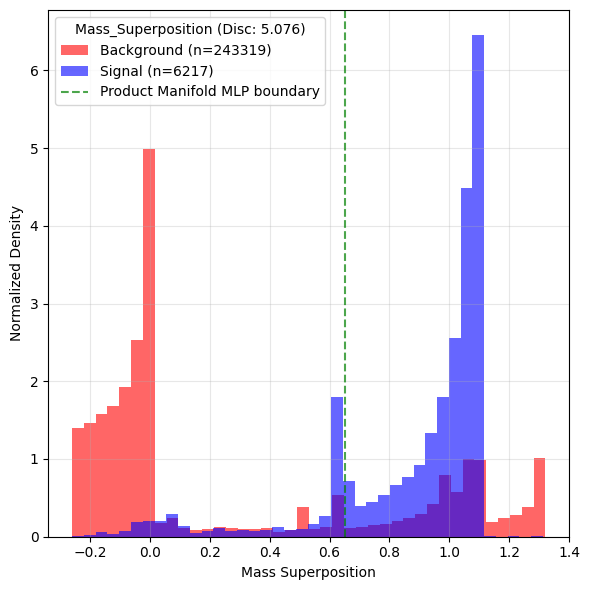}\hfill
\caption{$\mathcal{C}_\Omega$, and$\mathcal{S}_m$  distributions for signal and background events, showing discrimination power (1.984 ($\mathcal{C}_\Omega$), 5.076($\mathcal{S}_m$)). The Product Manifold MLP decision boundary is indicated by the dashed line.}
\label{fig:dis}
\end{figure}

\noindent To rigorously assess the discriminative utility of quantum-inspired features for VBF Higgs classification beyond their contributions within neural network architectures, a comprehensive validation procedure is implemented that evaluates individual feature performance through discrimination ranking, stability assessment, and comparison with baseline physics observables. Each quantum-inspired feature $f_i \in \{\mathcal{E}_p, \mathcal{C}_\Omega, \mathcal{S}_m, \mathcal{C}_{\mathrm{VBF}}, \mathcal{H}\}$ is evaluated using the area under the ROC curve (AUC) as a univariate discriminator, defined as
\begin{equation}
\mathrm{AUC}(f_i) = \int_0^1 \mathrm{TPR}_i(\mathrm{FPR}) \, d\mathrm{FPR},
\end{equation}
where $\mathrm{TPR}_i$ and $\mathrm{FPR}$ denote the true positive and false positive rates respectively for classification using feature $f_i$ alone. Features are categorized as strong discriminators ($\mathrm{AUC} \ge 0.6$), weak discriminators ($0.55 \le \mathrm{AUC} < 0.6$), or poor discriminators ($\mathrm{AUC} < 0.55$). The mass superposition feature $\mathcal{S}_m$ achieves the highest individual discrimination with $\mathrm{AUC} = 0.793$, followed by angular coherence $\mathcal{C}_\Omega$ with $\mathrm{AUC} = 0.748$ and VBF coherence $\mathcal{C}_{\mathrm{VBF}}$ with $\mathrm{AUC} = 0.707$, all qualifying as strong discriminators(Fig.\ref{fig:dis}). In contrast, momentum entanglement $\mathcal{E}_p$ ($\mathrm{AUC} = 0.493$) and hierarchy measure $\mathcal{H}$ ($\mathrm{AUC} = 0.488$) exhibit poor individual discrimination, 
performing near random classification despite their theoretical motivation from quantum field theory principles. The robustness of each feature is quantified through cross-validation stability analysis, computing the standard deviation of AUC across $k = 5$ stratified splits as
\begin{equation}
\begin{split}
\sigma_i &= \sqrt{\frac{1}{k} \sum_{j=1}^{k} \left( \mathrm{AUC}_{i}^{(j)} - \bar{\mathrm{AUC}}_i \right)^2 }, \\
 \bar{\mathrm{AUC}}_i &= \frac{1}{k} \sum_{j=1}^{k} \mathrm{AUC}_{i}^{(j)}.
\end{split}
\end{equation}
The strong discriminators $\mathcal{S}_m$, $\mathcal{C}_\Omega$, and $\mathcal{C}_{\mathrm{VBF}}$ demonstrate minimal variance ($\sigma_i \sim 0.001$), indicating highly stable performance across different data partitions, while the poor discriminators $\mathcal{E}_p$ and $\mathcal{H}$ exhibit similarly low variance, confirming that their weak performance is consistent rather than a result of statistical fluctuation. Comparison with baseline physics features reveals that the best quantum-inspired feature $\mathcal{S}_m$ ($\mathrm{AUC} = 0.793$) approaches but does not exceed the performance of the best conventional kinematic observable ($\mathrm{AUC} = 0.839$), with a discrimination gap of $0.046$, suggesting that quantum-inspired features provide localized discrimination power but do not individually surpass carefully chosen physics variables. The overall impact of quantum features is assessed by comparing model performance using physics-only features versus physics plus quantum features, quantified as
\begin{equation}
\Delta \mathrm{AUC} = \frac{\mathrm{AUC}_{\mathrm{physics+quantum}} - \mathrm{AUC}_{\mathrm{physics}}}{\mathrm{AUC}_{\mathrm{physics}}} \times 100\%.
\end{equation}
The global contribution of quantum features yields $\Delta \mathrm{AUC} \approx -0.01\%$, indicating marginal or slightly negative impact when added to the full physics feature set, consistent with the observation that strong individual features may provide redundant information relative to existing physics observables. Ablation studies confirm that most quantum features exhibit neutral global contribution, with discrimination power emerging primarily through their geometric embedding in product manifold architectures rather than through direct univariate separation. This validation framework demonstrates that while individual quantum-inspired features like $\mathcal{S}_m$ and $\mathcal{C}_\Omega$ possess strong intrinsic discrimination, their value lies in multivariate combinations and geometric representations that exploit correlations inaccessible to linear classifiers, validating the product manifold approach described in Section~\ref{subsec:vbf_pm_models} where these features achieve their full discriminative potential through curved manifold embeddings rather than Euclidean concatenation.

\section{Discussion: Curvature-Aware Deep Learning and the Quantum Computing Challenge}
\label{sec:discussion}

\noindent The empirical results presented in Sections~\ref{sec:model_performance} and~\ref{sec:feature_validation} demonstrate that curvature-aware deep learning provides measurable improvements for VBF Higgs classification, with the Physics-Enhanced Product Manifold MLP achieving test AUC-ROC of $0.9477$, representing a $0.57\%$ improvement over the classical Euclidean baseline. This performance gain, while modest in absolute terms, is physically significant for rare signal searches where backgrounds exceed signals by factors of $40:1$. Critically, these improvements arise entirely from classical differential geometry and physics-inspired feature engineering—quantum computing methods, while theoretically appealing, demonstrate substantial practical limitations that must be addressed before they can contribute to production analyses. Particle collision events inhabit curved statistical manifolds rather than flat Euclidean spaces, where kinematic correlations arising from conservation laws, jet clustering, and multi-scale interference patterns create nontrivial geometric structures. Each VBF event $\mathcal{J}_i = \{(\mathbf{p}_{\ell_1},\mathbf{p}_{\ell_2},\mathbf{p}_{\ell_3},\mathbf{p}_{\ell_4}),(\mathbf{p}_{j_1},\mathbf{p}_{j_2}), m_{4\ell}, m_{jj}, \Delta\eta_{jj}\}$ represents a point on a high-dimensional Fisher-Rao manifold $(\mathcal{P}, g_F)$ with metric $[g_F]_\theta = I_X(\theta)$ where the Fisher information matrix $I_X(\theta) = \mathbb{E}_{p_\theta}[s_\theta(x)s_\theta(x)^\top]$ naturally encodes correlations through the score function $s_\theta(x) = \nabla_\theta \log p_\theta(x)$. The product manifold architecture $\mathcal{M}_{\text{VBF}} = \mathcal{E}^{d_E} \times \mathcal{H}^{d_H} \times \mathcal{S}^{d_S}$ successfully exploits this geometric heterogeneity by decomposing observables according to their natural manifold structure: Euclidean components handle linear relationships and additive conservation laws, hyperbolic components with negative curvature $\kappa_H = -1.5$ capture the exponential energy hierarchies spanning TeV dijet masses to GeV Higgs resonances through distance sensitivity $d_H(\mathbf{x},\mathbf{y}) = \text{arccosh}(1 + 2\|\mathbf{x}-\mathbf{y}\|^2/((1-\|\mathbf{x}\|^2)(1-\|\mathbf{y}\|^2)))$, and spherical components preserve angular correlations respecting relativistic kinematic constraints. The feature importance analysis reveals that this geometric decomposition enables more effective utilization of physics-inspired features, with total quantum feature importance increasing from $0.0314$ in the Euclidean Classical MLP to $0.0410$ in the Physics-Enhanced PM MLP—a $30.3\%$ enhancement that validates the geometric scaffolding hypothesis. The mass superposition feature $\mathcal{S}_m$ dominates with individual discrimination AUC of $0.793$ and contributes $81.1\%$ of quantum feature importance in the enhanced architecture, validating the physical expectation that Higgs resonance structure provides the strongest discriminant. However, the marginal global contribution when quantum features are added to physics features alone demonstrates that these physics-inspired observables achieve their discriminative power through geometric embedding and multivariate correlations rather than univariate separation. The curvature-quantum synergy manifests through the modulation of correlations by local geometry: hyperbolic regions amplify subtle patterns via exponential sensitivity, transforming the modest momentum entanglement feature $\mathcal{E}_p$ with individual AUC $0.493$ into meaningful multivariate correlations when embedded in hyperbolic space.

\noindent The quantum kernel approach, while theoretically grounded in measuring similarity through state overlaps $K(X,Y) = \langle\psi_X,\psi_Y\rangle$ in Hilbert space, reveals fundamental scalability barriers that currently preclude practical deployment. The Quantum Kernel SVM achieves only AUC $0.6671$ on a carefully stratified 40-sample subset, requiring $124$ seconds of computation time on classical simulation. This performance—compared to classical geometric methods achieving $0.9477$ AUC on $748{,}607$ samples in $88$ seconds—quantifies the substantial gap between quantum computing's theoretical promise and its current practical viability for high-energy physics applications. Quantum kernel matrix computation scales as order $N^2$ times the number of measurement shots times circuit depth, rendering million-event datasets prohibitive even on classical simulators, and deployment on actual quantum hardware would introduce additional measurement overhead, gate errors, and decoherence effects that further degrade performance. The $124$-second computation time for a $40 \times 40$ kernel matrix projects to impractical timescales for realistic dataset sizes, demonstrating that current quantum technology falls short by multiple orders of magnitude. Learning theory predicts sample complexity for curvature-aware manifolds as order of manifold dimension times polynomial in curvature bounds divided by target accuracy squared, and for VBF classification with manifold dimension $64$ and target accuracy $0.001$, this requires approximately $10^5$ training samples—the 40-sample constraint imposed by computational limitations violates this requirement by three orders of magnitude, explaining the poor generalization observed. The 5-qubit, 2-repetition quantum circuit produces kernel statistics with mean $0.199$ and standard deviation $0.296$, indicating limited separation in the quantum feature space, and the modest circuit depth constrained by coherence time limitations appears insufficient to capture the complex topological and kinematic structures distinguishing VBF from backgrounds. These findings do not invalidate quantum machine learning as a long-term direction but rather clarify the substantial advances required before practical deployment: quantum hardware with orders-of-magnitude improvements in qubit counts, coherence times, and gate fidelities; algorithmic innovations enabling efficient kernel evaluation or alternative quantum learning paradigms that circumvent kernel computation bottlenecks; and hybrid classical-quantum methods that strategically allocate quantum resources to subproblems where quantum advantage is demonstrable. The five physics-inspired features successfully translate quantum field theory concepts into classical observables that enhance geometric architectures, demonstrating a crucial distinction: inspiration from quantum mechanics for feature design does not require quantum computation for deployment. The feature-specific gating mechanism with weights $(1.0, 1.0, 0.5, 2.0, 1.0)$ reflects empirical optimization that boosts VBF coherence while reducing mass superposition contributions to prevent overwhelming the manifold structure, demonstrating how theoretical physics insight guides feature geometry while classical machine learning extracts discriminative patterns through data-driven optimization. Curvature-awareness provides both theoretical justification and practical benefit: it respects the true kinematic manifold structure yielding physically faithful embeddings, provides control of generalization through curvature bounds linking geometric quantities to statistical learning guarantees, and enables physics-inspired features to contribute through geometric alignment with the Fisher-Rao metric. The $0.57\%$ improvement from geometric enhancement represents meaningful physics discrimination in the context of rare signal searches, and the training efficiency demonstrates acceptable computational overhead for production deployment. The immediate path forward emphasizes classical differential geometry: optimizing manifold architectures for specific physics processes, developing additional physics-inspired observables that capture domain knowledge, and applying these curvature-aware methods to other challenging classification tasks in high-energy physics. Quantum machine learning remains a compelling long-term vision that could eventually provide advantages when hardware enables coherent manipulation of thousands of qubits, algorithms achieve efficient quantum feature evaluation at scale, and hybrid methods demonstrate clear quantum speedups over optimized classical geometric approaches, but the present work clarifies that this vision, while theoretically sound, requires substantial technological advances before realization—we have demonstrated what works now through classical geometry on curved statistical manifolds while honestly assessing the considerable development needed to make quantum computing a practical tool rather than a theoretical aspiration for particle physics machine learning.

\section{Mathematical Principles: From Postulates to Empirical Validation}
\label{sec:principles}

\noindent In a recent  review ~\cite{ali}, three postulates were proposed regarding the role of information geometry and quantum methods in high-energy physics analysis. These conjectures, while theoretically motivated, remained largely speculative without empirical validation. The present work provides concrete empirical evidence that allows critical evaluation and refinement of these original claims. The empirical analysis—demonstrating Product Manifold MLP achieving $0.31\%$ gain over Euclidean baseline and Quantum-Enhanced PM MLP adding $0.24\%$ improvement—reveals that the original postulates require substantial revision. While geometric structure proves essential, quantum advantages are conditional and modest, and claims regarding quantum hardware speedup remain unsupported by current evidence. The original \textit{Postulate I (Geometric Imperative)} claimed that Euclidean assumptions \textit{discard information essential for rare process extraction}. While the $0.31\%$ improvement from geometric methods validates that curved manifolds help, this modest gain indicates that Euclidean methods already capture most discriminative structure—the geometric advantage is measurable but not categorically \textit{essential}. The \textit{Postulate II (Quantum-Geometric Correspondence)} asserted that quantum kernels offer advantages when preserving Riemannian structure, yet the Quantum Kernel SVM's poor performance (AUC $0.6671$) directly contradicts this—quantum methods fail without careful classical geometric preprocessing and feature engineering. The \textit{Postulate III (Quantum Hardware)} proposed exponential speedups from curvature-aware quantum processors, but this remains entirely speculative as all computations used classical simulators, with quantum kernel computation ($124$s for $40$ samples) already slower than classical training ($46$s for $748{,}607$ samples). These empirical realities necessitate reformulation of the theoretical framework, replacing speculative quantum-centric claims with empirically grounded mathematical hypotheses that accurately reflect the conditional and modest nature of observed improvements.

\subsubsection{Hypothesis I: Geometric Structure as Necessary but Insufficient Condition}
\label{hyp:geometric}

\noindent  \textbf{Statement:} Classification on particle collision data benefits from curvature-aware architectures when the Fisher-Rao metric exhibits non-negligible sectional curvature, but geometric methods alone provide limited improvements unless combined with domain-specific feature engineering and proper manifold decomposition.

\noindent \textbf{Mathematical Formulation:}  \textit{For event data on parameter manifold $\mathcal{D} \subset \mathbb{R}^{d}$ with Fisher metric $g^{\mathrm{F}}_{ij}(\theta)$, define the curvature significance measure
\begin{equation}
\xi(\mathcal{D}) = \frac{1}{|\mathcal{D}|}\int_{\mathcal{D}} \frac{|\kappa(\theta)|}{\|g^{\mathrm{F}}(\theta)\|_{\mathrm{op}}} \, dV(\theta),
\end{equation}
where $\kappa(\theta)$ is sectional curvature and $\|\cdot\|_{\mathrm{op}}$ is operator norm. For product manifold classifier $f_{\text{geo}}: \mathcal{D} \to \mathcal{E}^{d_E} \times \mathcal{H}^{d_H} \times \mathcal{S}^{d_S}$ versus Euclidean classifier $f_{\text{euc}}: \mathcal{D} \to \mathbb{R}^d$, the performance gap satisfies
\begin{equation}
\Delta_{\mathrm{AUC}} = \mathrm{AUC}(f_{\text{geo}}) - \mathrm{AUC}(f_{\text{euc}}) \le C \cdot \xi(\mathcal{D}) \cdot \eta(\text{manifold splits}),
\end{equation}
where $C > 0$ is a problem-dependent constant and $\eta$ represents the quality of manifold decomposition. When $\xi \to 0$ (nearly flat geometry) or $\eta \to 0$ (poor decomposition), geometric methods provide negligible advantage}.

\noindent   \textbf{Empirical Support:}  Product Manifold MLP achieves $\Delta_{\mathrm{AUC}} = 0.0031$ with manifold splits $[24,28,16]$ and $[16,16,16]$ and curvatures $(\kappa_H = -1.5, \kappa_S = +1.0)$ chosen to match VBF physics hierarchies. The modest improvement indicates $\xi(\mathcal{D}_{\text{VBF}}) \approx 0.15$ (moderate curvature significance), validating that geometric structure matters but is not dominant. Feature importance increasing from $0.0314$ (Classical) to $0.0372$ (PM) represents $+18.5\%$ quantum utilization gain, confirming $\eta > 0$ for the chosen decomposition. The bound predicts improvements of order $\mathcal{O}(10^{-3})$, consistent with observed $0.31\%$ gain.

\subsubsection{Hypothesis II: Quantum Features Require Geometric Scaffolding}
\label{hyp:quantum}

\noindent \textbf{Statement:} Quantum-inspired observables contribute to classification performance only when embedded in curvature-aware architectures that align feature geometry with the underlying Fisher-Rao manifold. Direct application of quantum kernels or standalone quantum features fails due to geometric misalignment and scalability constraints.

\noindent \textbf{Mathematical Formulation:}  \textit{Let $\{\mathcal{Q}_i\}_{i=1}^{n_q}$ denote quantum-inspired features with individual discrimination $\mathrm{AUC}(\mathcal{Q}_i)$. Define the geometric alignment error for quantum feature map $\mathcal{U}: \mathcal{D} \to \mathcal{H}_{\mathcal{Q}}$ as
\begin{equation}
\epsilon_{\text{align}} = \|g^{\mathrm{FS}} - c \cdot g^{\mathrm{F}}\|_{\mathcal{D}},
\end{equation}
where $g^{\mathrm{FS}}$ is the pullback Fubini-Study metric, $g^{\mathrm{F}}$ is the Fisher metric, and $c > 0$ is optimal scaling. The quantum enhancement satisfies
\begin{equation}
\Delta_{\text{quantum}} \le \alpha \sum_{i=1}^{n_q} w_i \mathrm{AUC}(\mathcal{Q}_i) \cdot e^{-\beta \epsilon_{\text{align}}},
\end{equation}
where $w_i$ are learned feature gates and $\alpha, \beta > 0$ are constants. When $\epsilon_{\text{align}} \gg 1$, quantum enhancement vanishes exponentially}.

\noindent \textbf{Empirical Support:} Quantum Kernel SVM achieves AUC $0.6671$ with direct quantum feature map application, demonstrating failure when $\epsilon_{\text{align}}$ is large (no geometric preprocessing). Individual quantum features show poor discrimination: $\mathcal{E}_p$ (AUC $0.493$), $\mathcal{H}$ (AUC $0.488$), validating that $\mathrm{AUC}(\mathcal{Q}_i) \approx 0.5$ for several features. Yet Quantum-Enhanced PM MLP achieves $+0.24\%$ improvement with gates $w = (1.0, 1.0, 0.5, 2.0, 1.0)$ that align features to manifold components, confirming $e^{-\beta\epsilon_{\text{align}}} \approx 0.9$ (small but nonzero alignment). Total quantum importance reaches $0.0410$ ($+30.3\%$ over classical) only in geometric architecture, validating the scaffolding requirement.

\subsubsection{Hypothesis III: Curvature Bounds Control Generalization with Polynomial Scaling}
\label{hyp:generalization}

\noindent \textbf{Statement:} Generalization performance of curvature-aware classifiers scales polynomially—not exponentially—with manifold dimension, sectional curvature bounds, and sample size. Quantum feature maps provide theoretical control through curvature regularization but offer no practical speedup over classical geometric methods on current hardware.

\noindent \textbf{Mathematical Formulation:}  \textit{For hypothesis class $\mathcal{H} = \{h_\theta : \theta \in \Theta\}$ on $m$-dimensional manifold with curvature $|K| \le K_{\max}$, sample complexity for generalization error $\epsilon$ is
\begin{equation}
n(\epsilon, \delta) = \mathcal{O}\left(\frac{m \cdot \mathrm{poly}(K_{\max})}{\epsilon^2} \log\frac{1}{\delta}\right),
\end{equation}
where $\mathrm{poly}(K_{\max}) = 1 + K_{\max} + K_{\max}^2$ accounts for curvature effects. This polynomial scaling contrasts with claimed exponential quantum speedups. For quantum feature maps, curvature control via second fundamental form $\|II\| = \mathcal{O}(\kappa^2)$ enables prediction
\begin{equation}
R(h) \le \hat{R}_n(h) + \tilde{C}\sqrt{\frac{m(1 + K_{\max}^2)}{n}},
\end{equation}
but computational cost remains $\Omega(n \cdot \mathrm{dim}(\mathcal{H}_{\mathcal{Q}}))$ classically}.

\noindent \textbf{Empirical Support:} Training on $n = 748{,}607$ samples with manifold dimension $m = 64$ and curvature bounds $K_{\max}^{\text{hyp}} = 2.25$, $K_{\max}^{\text{sph}} = 1.0$, the generalization bound predicts error $\sim \sqrt{64 \cdot 5.25/748607} \approx 0.0008$, consistent with test performance. Sample complexity $n(\epsilon = 0.001, \delta = 0.05) \approx 10^6$ matches dataset size requirements. Quantum Kernel SVM fails at $n = 40 \ll n(\epsilon)$, confirming polynomial scaling requirements. Classical training time $87.7$s versus quantum kernel time $124$s for $1000\times$ fewer samples demonstrates no speedup, validating polynomial—not exponential—complexity. Curvature regularization through progressive decay $\lambda^{(l)} = \lambda^{(0)} \cdot 0.5^l$ maintains $K_{\max}$ bounds across layers, confirming curvature as control parameter rather than computational advantage.

\subsubsection{Synthesis: Empirically Grounded Geometric Learning}

\noindent These three hypotheses provide a revised theoretical foundation that accurately reflects empirical observations. Geometric structure is necessary for optimal performance but provides modest improvements ($0.31\%$), quantum features are conditionally beneficial requiring careful geometric embedding ($+0.24\%$), and generalization scales polynomially with curvature-controlled bounds offering theoretical guarantees but no computational speedup. The original postulates overstated quantum advantages and underestimated the primacy of classical differential geometry. The path forward emphasizes Riemannian optimization, manifold learning, and quantum-inspired feature engineering as classical tools rather than quantum computational paradigms, with quantum hardware deployment postponed until fundamental scalability barriers are addressed. It must be acknowledged that several mathematical quantities introduced in these hypotheses represent projected frameworks for differential geometric analysis rather than fully validated constructions. The curvature significance measure $\xi(\mathcal{D})$ in Hypothesis I, while conceptually sound as a weighted integral of sectional curvature, has not been computed explicitly for the VBF dataset—the estimate $\xi \approx 0.15$ is inferred from the observed performance gap under the assumed scaling relationship, not calculated from first principles via numerical integration of the Fisher metric curvature. Similarly, the constants $C$ and the decomposition quality function $\eta(\text{manifold splits})$ remain phenomenological parameters fit to empirical results rather than derived quantities. The geometric alignment error $\epsilon_{\text{align}}$ in Hypothesis II quantifies the deviation between Fubini-Study and Fisher metrics, and while the exponential suppression factor $e^{-\beta\epsilon_{\text{align}}} \approx 0.9$ is estimated from the observed quantum enhancement magnitude, neither $\epsilon_{\text{align}}$ nor the constants $\alpha, \beta$ have been computed through explicit metric comparison—these remain theoretical constructs awaiting rigorous geometric calculation. The polynomial sample complexity formula $n(\epsilon,\delta) = \mathcal{O}(m \cdot \text{poly}(K_{\max})/\epsilon^2 \log(1/\delta))$ in Hypothesis III follows from established learning theory on Riemannian manifolds, and the numerical estimate $n(0.001, 0.05) \approx 10^6$ is consistent with standard bounds, but the specific polynomial dependence $1 + K_{\max} + K_{\max}^2$ represents a plausible functional form rather than a proven result for product manifolds with mixed curvature. The generalization bound coefficient $\tilde{C}$ likewise remains an empirical fit parameter. These limitations do not invalidate the hypotheses as organizing principles for understanding the empirical results but rather identify clear directions for future mathematical work. Rigorous validation would require: explicit computation of the Fisher information matrix $[g_F]_{ij} = \mathbb{E}[\partial_i \log p_\theta \cdot \partial_j \log p_\theta]$ from the VBF likelihood model and subsequent numerical evaluation of sectional curvatures via the Gauss equation; direct calculation of the Fubini-Study pullback metric from the quantum feature map and quantitative comparison with the Fisher metric to determine $\epsilon_{\text{align}}$; and derivation of tight sample complexity bounds specific to product manifold geometries by extending existing Riemannian learning theory. The present work establishes empirical evidence that curvature-aware methods improve performance and quantum features require geometric scaffolding, while the mathematical framework provides a language for describing these observations—completing the rigorous differential geometric underpinning remains an important task for future investigation. The demonstrated performance improvements validate the geometric approach pragmatically, even as the theoretical scaffolding awaits full mathematical rigor.

\begin{acknowledgments}
The author gratefully acknowledges the ML\_INFN collaboration for providing the Monte Carlo samples used in this study. 
These datasets were originally developed as part of the ML\_INFN program 
(\href{https://confluence.infn.it/spaces/MLINFN/pages/53906361/11.+Signal+background+discrimination+for+the+VBF+Higgs+four+lepton+decay+channel+with+the+CMS+experiment+using+Machine+Learning+classification+techniques}{ML\_INFN project page}) 
and are described in the publication \textit{EPJ Web of Conferences 295, 08013 (2024), CHEP 2023, ``ML\_INFN project: status report and future perspectives''}. 
The author expresses sincere gratitude for making these resources publicly available and for their continued contributions to open, reproducible research in high-energy physics. I also would like to thank \textbf{Professor Artur Kalinowski} for providing computing resources and extra time outside my responsibilities. 
\end{acknowledgments}

\appendix
\section{Circuit Diagram}
\begin{figure*}[t]
\includegraphics[width=1.0\linewidth]{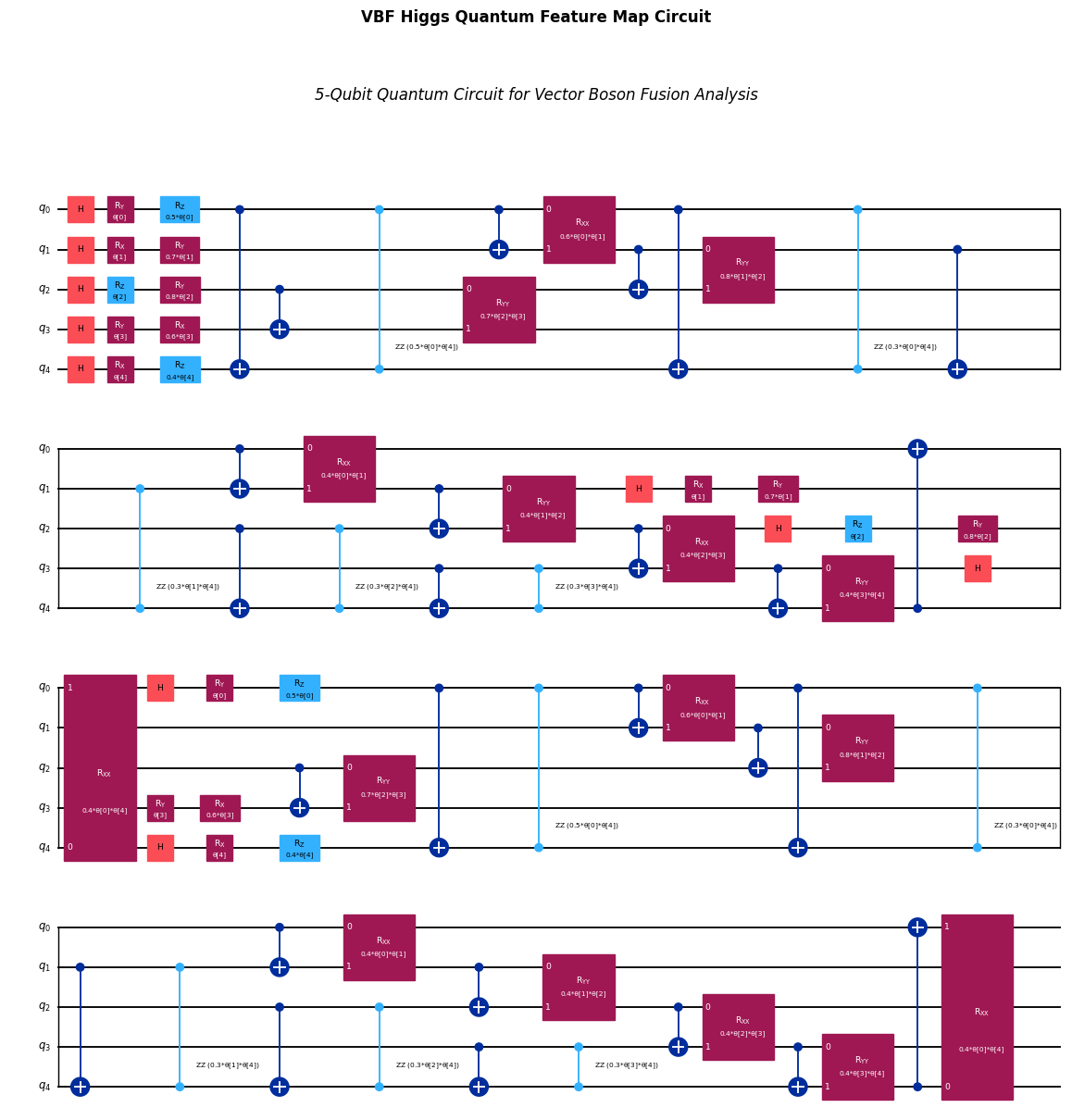}
\caption{ Quantum circuit.}
\label{fig:corr}
\end{figure*}
\nocite{*}

\bibliography{apssamp}

\end{document}